\documentclass[aps, pra, reprint, superscriptaddress, longbibliography]{revtex4-2}

\usepackage{graphicx}

\usepackage{amsmath, amssymb}
\usepackage[utf8]{inputenc}
\usepackage{braket}

\usepackage{color}
\usepackage{hyperref}

\begin{document}

\title{Nonlocal Magnonic Cat States in Hybrid Magnon-Qubit Architectures}



\author{Urmimala Dewan}
\thanks{Corresponding author: d.urmimala@iitg.ac.in}
\affiliation{Department of Physics, Indian Institute of Technology Guwahati, Guwahati-781039, India}

\author{Roson Nongthombam}
\affiliation{Department of Physics, Indian Institute of Technology Guwahati, Guwahati-781039, India}

\author{Sampreet Kalita}
\affiliation{Department of Physics, University of Strathclyde, Glasgow G4 0NG, United Kingdom}

\author{Amarendra~K.~Sarma}
\affiliation{Department of Physics, Indian Institute of Technology Guwahati, Guwahati-781039, India}

\date{\today}

\begin{abstract}
The quantum superpositions of coherent states offer an alternative to the conventional qubit-based encodings by harnessing the large Hilbert space available in bosonic modes, including those realised in microwave and optical cavities, magnons, and mechanical resonators. 
Beyond their advantages for local information processing, establishing long-distance quantum networks for such bosonic states is crucial for scalable quantum communication and distributed quantum computation.
In this work, we propose an entanglement-swapping-based protocol to generate a bipartite magnonic cat state shared between spatially separated subsystems. Each subsystem comprises a hybrid architecture consisting of a superconducting transmon qubit coupled to a yttrium iron garnet (YIG) sphere that supports magnon modes. By performing a projective Bell-state measurement on the qubits, the initially established magnon-qubit entanglement is coherently transferred to the remote magnon modes, resulting in a nonlocal magnonic cat state. 
For experimental characterisation of the generated states, we perform quantum state tomography through reconstruction of the Wigner function using joint displaced parity measurements of the magnon modes.
Our scheme provides a feasible route towards realising long-distance magnonic entanglement and contributes to the advancement of hybrid quantum network architectures. 
\end{abstract}

\maketitle

\section{Introduction}
\label{sec:introduction}

Long-distance quantum networks form the backbone of scalable quantum communication and large-scale quantum information processing~\cite{PhysRevLett.81.5932,Kimble2008,doi:10.1126/science.aam9288}.
An essential milestone toward realising such quantum networks is the ability to prepare and manipulate nonclassical states \cite{Deleglise2008,PhysRevA.64.022313,PhysRevLett.65.3385,PhysRevA.56.4175}.
Of particular interest is the Schrödinger cat state -- a macroscopic superposition of coherent states -- realised in systems such as optical \cite{hacker.13.110,doi:10.1126/science.1243289}, superconducting qubits \cite{PhysRevA.109.023703}, mechanical resonators \cite{doi:10.1126/science.adf7553,hou2016generation,EtehadiAbari:20}, and magnons \cite{zhgm-p3ss, PhysRevLett.127.087203}.
Such nonclassical states are crucial for quantum metrology \cite{PhysRevLett.107.083601,Nature.535.262} and for exploring the quantum-to-classical transition \cite{PhysRevLett.77.4887}, while also being highly relevant to fault-tolerant quantum computing \cite{PhysRevA.68.042319,PhysRevLett.100.030503,PhysRevA.65.042305}, quantum communication \cite{PhysRevA.78.062319,9024573}, and quantum simulation \cite{wbp6-y3vd}.
By entangling them with another continuous-variable mode, one can realise bipartite cat states that manifest quantum entanglement of coherent states \cite{Sanders_2012, SciAdv.8.eabn1778}.
The creation of bipartite cat states in photonic \cite{Science.352.1087,SciAdv.8.eabn1778} and phononic \cite{PhysRevA.110.023726} modes has been studied in previous works.
Beyond their scientific exploration of nonlocality \cite{PhysRevA.67.012105}, bipartite cat states enable entanglement distribution, state transfer, and precision sensing in quantum networks. 

Systems based on magnons, collective spin excitations in magnetic materials, are particularly attractive in this context owing to their long lifetimes and high spin densities \cite{doi:10.1126/sciadv.1501286, 2021NatRM...6.1114P} and their ability to couple to diverse degrees of freedom, including photons, phonons, and superconducting qubits \cite{PhysRevLett.113.156401,PhysRevLett.113.083603,PhysRevLett.121.203601,PhysRevLett.125.117701,doi:10.1126/science.aaz9236}.
YIG-sphere-based magnonic systems, distinguished by their tunability and low dissipation, and compatibility with other hybrid systems harnessing their distinct advantages have found broad applications in quantum technologies. 
In recent years, magnonic systems have demonstrated significant potential for the development of quantum networks \cite{PRXQuantum.2.040344,PhysRevA.100.022343}. 
Experimental demonstration of magnon-qubit interaction through virtual photons of microwave cavity was shown in \cite{doi:10.1126/science.aaa3693} and direct coupling of magnon-qubit via magnetic flux has also been theoretically proposed \cite{PhysRevLett.129.037205}. Furthermore, various protocols for generating magnonic cat states have already been explored \cite{PhysRevA.110.013711,PhysRevApplied.21.044018,PhysRevA.112.023709} in magnon-qubit hybrid architecture. 

In this work, we propose to create magnonic bipartite cat states using an entanglement-swapping scheme between a spatially separated hybrid system pair.
Specifically, we consider direct coupling between a magnon and a superconducting qubit via magnetic flux, thereby generating entanglement between them.
The qubits in the hybrid system pairs interact through a microwave cavity, resulting in the generation of qubit Bell states.
After performing a projective measurement \cite{PhysRevLett.123.060502} on the qubit Bell states, the magnonic modes become entangled, thereby enabling the remote generation of four magnonic bipartite cat states.
Recent studies have explored magnon-magnon entanglement in a variety of hybrid quantum platforms, such as optomagnonic systems with optical Bell-state measurements \cite{Golkar_2024}, antiferromagnetic sublattice coupling quantified via microwave cavities \cite{PhysRevB.104.224302}, microwave cavity architectures where superconducting qubits mediate long-range magnon interactions \cite{PhysRevB.105.094422}, and spinning whispering-gallery-mode cavities \cite{xu2026nonreciprocal}.
In contrast to these works, in which entanglement is established directly between magnon modes, our scheme generates a nonlocal magnonic cat state by transferring magnon-qubit entanglement to remote magnon modes via an entanglement-swapping protocol.
Such entanglement-swapping schemes play a crucial role in quantum repeaters, which are essential for the realisation of long-distance quantum communication. 

To visualise and fully characterise the generated entangled states, we employ quantum state tomography based on Wigner function reconstruction via joint parity measurements.
The theoretical connection between parity and Wigner function reconstruction was clarified by Lutterbach and Davidovich \cite{PhysRevLett.78.2547}, and experimentally realised in the optical domain \cite{PhysRevLett.89.200402,PhysRevA.60.674}.
In circuit quantum electrodynamics, the dispersive interaction naturally enables quantum non-demolition (QND) readout of parity, which has been widely employed to reconstruct single-mode Wigner functions of nonclassical states with high fidelity \cite{Nature.511.444}. 
For bipartite systems, this relation generalises to joint parity measurements, where simultaneous access to both modes is required to reconstruct the joint Wigner function and capture intermode quantum correlations.
Such joint Wigner tomography has been experimentally realised in 
qubit-cavity systems \cite{Vlastakis2015} and in coupled two-cavity 
architectures, where it was used to reconstruct two-mode cat states 
\cite{Science.352.1087}. 
This protocol directly extends to hybrid magnonic systems, where bosonic magnon modes dispersively coupled to a superconducting qubit enable QND measurement of joint magnon-number parity, forming the basis for joint Wigner tomography.
Our generation scheme is convenient to implement, as the system features a distinct and tunable coupling between the magnon and the superconducting circuit. This coupling can be conveniently switched on and off via flux modulation.

Our paper is arranged as follows.
Sec. \ref{sec:system_hamiltonian} describes the magnon-transmon hybrid system, followed by Sec. \ref{sec:entanglement}, which outlines the generation of magnon-qubit entanglement.
The procedure for generating bipartite magnonic cat states is presented in Sec. \ref{sec:bipartite}.
Sec. \ref{sec:tomography} is devoted to the physical realisation and characterisation of these cat states through joint quantum state tomography.


\section{The System Hamiltonian}
\label{sec:system_hamiltonian}

\begin{figure}[tp]
    \includegraphics[width=.98\linewidth]{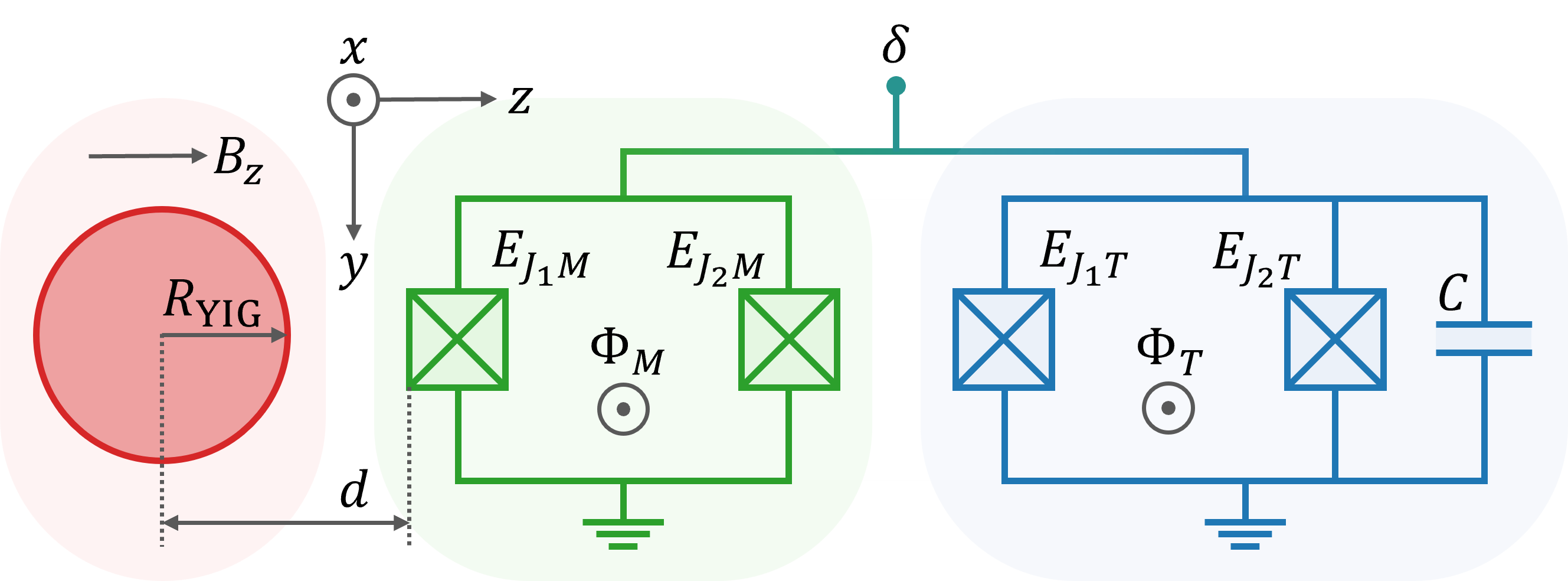}
    \caption{
        (color online) Schematic representation of the hybrid magnon-qubit system consisting of a YIG sphere (left, red) connected to a transmon SQUID (right, blue) via a magnon-coupling SQUID (center, green).
    }
    \label{fig:schematics}
\end{figure}

We consider a hybrid quantum system consisting of a ferromagnetic yttrium iron garnet (YIG) sphere supporting magnetic excitations (magnons), which is coupled to a superconducting qubit via the flux induced by the sphere.
The qubit is a flux-tunable transmon formed by a symmetric superconducting quantum interference device (SQUID) with Josephson energies $E_{J_{1} T}$ and $E_{J_{2} T}$, and shunted by a large capacitor $C$.
To introduce a tunable coupling between the transmon qubit and the magnetostatic mode of the YIG sphere (magnon), a second SQUID with Josephson energies $E_{J_{1} M}$ and $E_{J_{2} M}$ is connected in parallel to it.
The schematic of this setup is shown in Fig. \ref{fig:schematics}.
The transmon SQUID and the magnon-coupling SQUID are independently threaded by external fluxes $\Phi_{T}$ and $\Phi_{M}$, respectively.
In general, tunable couplers possess an external control parameter that allows the effective coupling to be turned on or off.
Here, by controlling $\Phi_{M}$, the effective coupling between the magnon-coupling SQUID and the magnon mode can be dynamically adjusted, enabling continuous tuning and complete suppression whenever required. 
The Hamiltonian for the effective transmon circuit is given by \cite{PhysRevLett.129.037205}
\begin{align}
\label{eqn:hamiltonian_squids}
{\hat{H}_{TM}^{(0)}} = 4 E_{c} \hat{n}^{2} 
- \bigg[ & E_{JM}^{\mathrm{max}} \cos\left( \frac{\pi \Phi_{M}}{\Phi_{0}} \right) \nonumber \\
& + E_{JT}^{\mathrm{max}} \cos\left(\frac{\pi \Phi_{T}}{\Phi_{0}}\right) \bigg] \cos{\hat{\delta}}.
\end{align}

Here, $E_{c}$ is the charging energy, and $\hat{n}$ is the number operator corresponding to tunnelling Cooper pairs.
Its canonically conjugate variable is the superconducting phase difference $\hat{\delta} = 2 \pi \hat{\Phi} / \Phi_0$.
Furthermore, $E_{JT}^{\mathrm{max}} = E_{J_{1} T} + E_{J_{2} T}$ and $E_{JM}^{\mathrm{max}} = E_{J_{1} M} + E_{J_{2} M}$ denote the maximum Josephson energies of the transmon SQUID and the magnon-coupling SQUID respectively.

An in-plane magnetic field $B_z$ is applied to the YIG sphere.
The magnetisation induced by this field produces a fluctuating magnetic moment $\Delta \mu$, which in turn generates a magnetic flux in the magnon-coupling SQUID.
This flux modulates the SQUID's inductive energy and shifts its frequency through the effective flux $\Phi_{M} + \Phi(\Delta \mu)$.
Here, the induced magnetic flux $\Phi(\Delta \mu)$ introduces a direct coupling between the magnon and the transmon without the need for a mediating microwave cavity \cite{PhysRevLett.129.037205}.

The fundamental magnetostatic mode of the sphere is the Kittel mode, which corresponds to uniform precession of the magnetisation.
The Hamiltonian of this Kittel mode can be written as $\hat{H}_{0} = \hslash \omega_{m} \hat{m}^{\dagger} \hat{m}$, where $\omega_{m} = \eta B_{z}$ is the ferromagnetic resonance frequency, $\eta$ is gyromagnetic ratio and $\hat{m}^{\dagger}$ ($\hat{m}$) is the magnon creation (annihilation) operator \cite{PhysRev.58.1098}.
Assuming $\Phi(\Delta \mu) \ll \Phi_{0}$, the total Hamiltonian of the hybrid magnon-qubit system can then be written as
\begin{align}
\label{eqn:hamiltonian_total_approx}
\hat{H} & = \hslash \omega_{m} \hat{m}^{\dagger} \hat{m} + E_{JM}^{\mathrm{max}} \sin{\left( \frac{\pi \Phi_{M}}{\Phi_{0}} \right)} \Phi(\Delta \mu) \cos{\hat{\delta}} + 4 E_{c} \hat{n}^{2} \nonumber \\
& - \bigg[ E_{JM}^{\mathrm{max}} \cos{\left( \frac{\pi \Phi_{M}}{\Phi_0} \right)} + E_{JT}^{\mathrm{max}} \cos{\left( \frac{\pi \Phi_{T}}{\Phi_{0}} \right)} \bigg] \cos{\hat{\delta}}.
\end{align}

The last three terms in Eq. \eqref{eqn:hamiltonian_total_approx} describe the effective transmon.
In terms of the bosonic creation and annihilation operators, the effective Hamiltonian of the transmon-magnon coupled system under rotating-wave approximation (RWA) can be approximated to (refer to Appendix \ref{app:hamiltonian} for its derivation)
\begin{equation}
\label{eqn:hamiltonian_rwa}
\hat{H} = \hslash \omega_{m} \hat{m}^{\dagger} \hat{m} + \hslash \tilde{g} \left( \hat{m}^{\dagger} + \hat{m} \right) \hat{c}^{\dagger} \hat{c} + \hslash \omega_{q} \hat{c}^{\dagger} \hat{c} - \frac{E_{c}}{2} \hat{c}^{\dagger} \hat{c}^{\dagger} \hat{c} \hat{c}.
\end{equation}

Here, the transmon frequency is $\omega_{q} = \sqrt{8 E_{c} A} - E_{c}$, where $A = E_{JM}^{\mathrm{max}} \cos{\left( \pi \Phi_{M} / \Phi_{0} \right)}+ E_{JT}^{\mathrm{max}} \cos{\left( \pi \Phi_{T} / \Phi_{0} \right)}$
The second term represents the nonlinear magnon-transmon coupling with strength
\begin{equation}
\label{eqn:coupling_strength}
\tilde{g} = \frac{E_{JM}^{\mathrm{max}} \mu_{0} \mu_{\text{ZPF}} \sqrt{2 E_{c}}}{4 \hslash \Phi_{0} d_{\text{min}} \sqrt{A}}
\sin{\left( \frac{\pi \Phi_{M}}{\Phi_{0}} \right)},
\end{equation}
where $\mu_{0}$ is vacuum permeability, $\mu_{\mathrm{ZPF}} = \gamma \sqrt{N_s / 2}$ is the zero-point fluctuation of the magnetisation and $d_{\mathrm{min}}$ is the shortest distance between the YIG sphere and the coupler SQUID, given by $d_{\mathrm{min}} =\sqrt{2} R_{\mathrm{YIG}}$, $R_{\mathrm{YIG}}$ being the radius of the YIG sphere and $N_{s} = 2 \times 10^{12}$ the total number of spins \cite{doi:10.1126/science.aaa3693}.

To enhance the nonlinearity, a time-dependent modulation of the magnon-transmon coupling is introduced by parametrically modulating the flux of the magnon-coupling SQUID through a local flux line \cite{PhysRevApplied.6.064007}.
A weak ac flux bias $\Phi_{M} = \Phi_{\mathrm{ac}} \cos(\omega_{\mathrm{ac}} t)$ with $\pi \Phi_{\mathrm{ac}} \ll \Phi_{0}$ is applied.
Substituting this into Eq. \eqref{eqn:coupling_strength}, the final Hamiltonian in the magnon rotating frame (given by the unitary $\hat{U}_{m} = e^{i \omega_m t \hat{m}^{\dagger} \hat{m}}$) under RWA becomes (refer to Appendix \ref{app:modulation} for its derivation)
\begin{equation}
\label{eqn:hamiltonian_final}
\hat{H} = \hslash \Delta \hat{m}^{\dagger} \hat{m}+ \hslash g (\hat{m}^{\dagger} + \hat{m}) \hat{c}^{\dagger} \hat{c} + \hslash \omega_{q} \hat{c}^{\dagger} \hat{c} - \frac{E_{c}}{2} \hat{c}^{\dagger} \hat{c}^{\dagger} \hat{c} \hat{c},
\end{equation}
where $\Delta = \omega_{m} - \omega_{\mathrm{ac}}$ and $g = 2 \Gamma$ are the effective detuning and effective magnon-transmon coupling with the normalisation constant
\begin{equation}
\label{eqn:Gamma}
\Gamma = \frac{E_{JM}^{\mathrm{max}} \Phi_{\mathrm{ac}} \mu_{0} \mu_{\text{ZPF}} \sqrt{2 E_{c}}}{16 \hslash \Phi_{0} d_{\text{min}} \sqrt{E_{JM}^{\mathrm{max}} + E_{JT}^{\mathrm{max}} \cos{\left( \pi \Phi_{T} / \Phi_{0} \right)}}}.
\end{equation}

Using standard parameters for a transmon qubit and a YIG sphere \cite{PhysRevLett.129.037205}, $d_{\mathrm{min}} = 4.2 \mu$m, $E_{c} / \hslash = 200$ MHz, $E_{JM}^{\mathrm{max}} / \hslash = 30$ GHz, $E_{JT}^{\mathrm{max}} / \hslash = 60$ GHz, $\Phi_{\mathrm{ac}} / \Phi_{0} = 0.1$, one obtains $\Gamma \approx 1$ MHz.
For our numerical analysis, we simulate our systems using QuTiP \cite{PhysRep.1153.1} in units of the normalisation constant $\Gamma$.


\section{Generation of the Magnon-Qubit Entangled State}
\label{sec:entanglement}

The transmon-qubit is prepared in the superposition of its ground ($\ket{0}$) and excited ($\ket{1}$) states, i.e. $\ket{+} = (\ket{0} + \ket{1}) / \sqrt{2}$.
In the transmon regime ($E_{c} \ll (E_{JM}^{\mathrm{max}}+ E_{JT}^{\mathrm{max}})$) and under resonant modulation ($\Delta = 0$), the dynamics is predominantly governed by the non-linear interaction term $\hat{H}_{\mathrm{int}} = \hslash g \left( \hat{m}^{\dagger} + \hat{m} \right) \hat{c}^{\dagger} \hat{c}$. 
This results in qubit-state-dependent coherent magnon displacement.
The resultant state is $\ket{\psi} = ( \ket{0}_{m} \ket{0}_{q} + \ket{\beta}_{m}\ket{1}_{q} ) / \sqrt{2}$, where $\beta = -i g t$.
This state corresponds to an entangled Bell-cat state between the qubit and the magnon modes.

Dissipative dynamics of this evolved state is incorporated by the Lindblad master equation
\begin{equation}
\label{eqn:lindblad_evolution}
\dot{\rho}(t) = - \frac{i}{\hslash}[\hat{H}_{\mathrm{int}}, \rho] + \gamma_{m} \mathcal{L}[\hat{m}] \rho + \gamma_{q} \mathcal{L}[\hat{c}] \rho,
\end{equation}
where, $\mathcal{L}(\rho) = \sum_{k} (\mathcal{O}_{k} \rho \mathcal{O}_k^{\dagger}
- \frac{1}{2} \{\mathcal{O}_{k}^{\dagger} \mathcal{O}_{k} , \rho\})$ is the Lindblad superoperator describing the dissipation channels $\{\mathcal{O}_{k}\}$. 
Here, $\gamma_{m}$ is the magnon decay rate and $\gamma_{q}$ is the qubit relaxation rate.

Under the dissipative dynamics of the system, the target state (denoted by the density matrix $\rho_{\mathrm{target}} = \ket \psi \bra \psi$) gets degraded and the similarity between the degraded state (denoted by $\rho$) and the target state is determined using the Uhlmann fidelity, defined as $F(\rho, \rho_{\mathrm{target}}) = ( \mathrm{Tr} [ \sqrt{\sqrt{\rho} \rho_{\mathrm{target}} \sqrt{\rho}} ] )^{2}$.
To quantify the degree of entanglement between magnon and qubit, we use the logarithmic negativity, defined as $E_{N} (\rho) = \log_{2} \vert \vert \rho^{\mathrm{T}_{p}} \vert \vert$, where $\vert \vert \rho^{T_{p}} \vert \vert$ is the trace norm of the partial transpose ($\mathrm{T}_{p}$) of the bipartite mixed state $\rho$ \cite{PhysRevA.65.032314}.

\begin{figure}[tp]
    \centering
    \includegraphics[width=.96\linewidth]{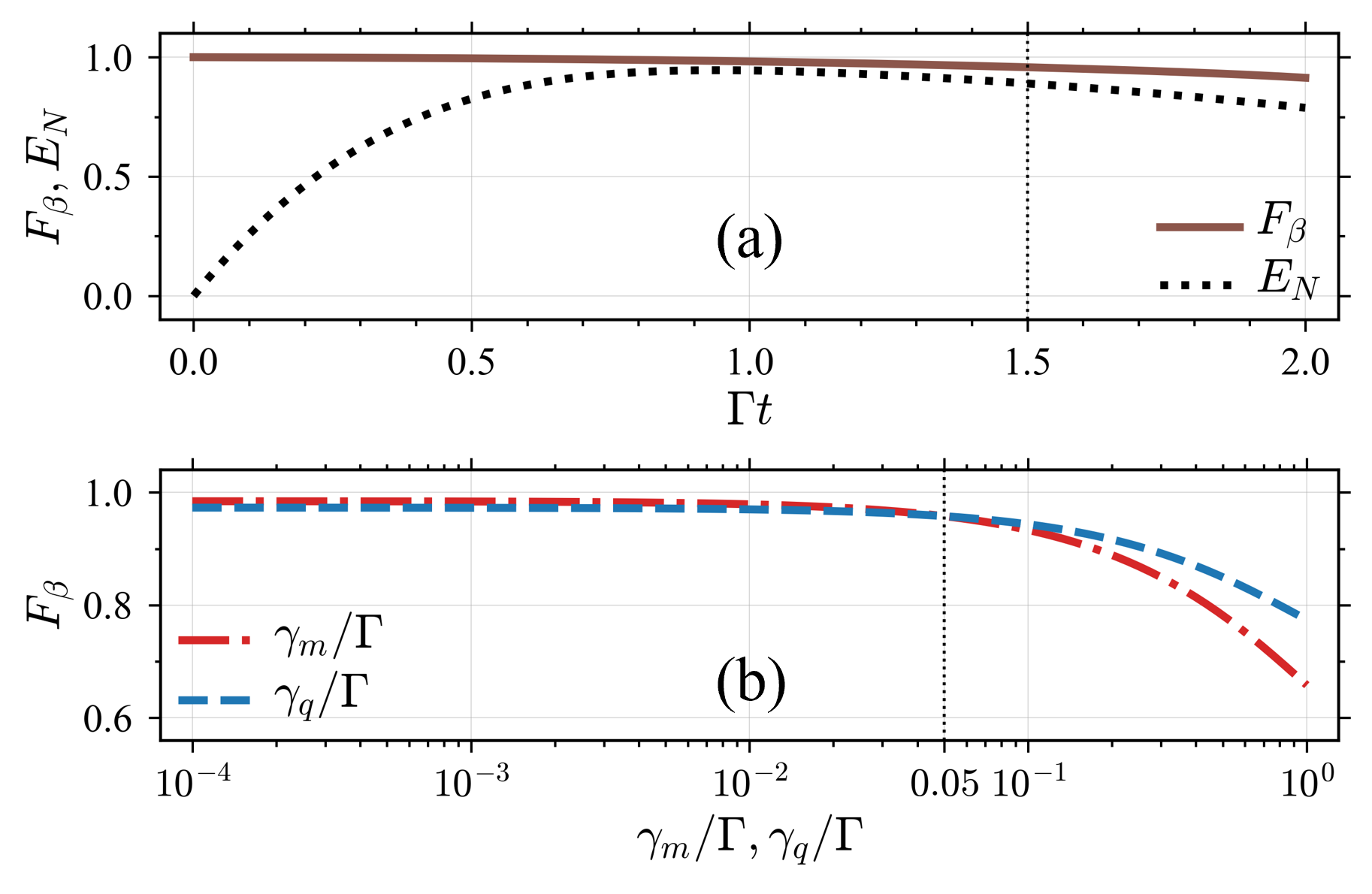}
    \caption{
        (color online) (a) Fidelity $F_{\beta}$ (solid brown) between the magnon-qubit state obtained using Eq. \eqref{eqn:lindblad_evolution} and the target state $\ket{\psi} = ( \ket{0}_m\ket{0}_{q} + \ket{\beta (t)}_m \ket{1}_{q})/ \sqrt{2}$, with $\beta(t) = -igt$, and corresponding magnon-qubit entanglement $E_{N}$ (dotted black) in the evolved state.
        (b) Variation of fidelity with the decay rates of the magnon (dash-dotted red) and the qubit (dashed blue) at $t = 1.5 / \Gamma$.
        Here, $\gamma_{m}, \gamma_{q} = 0.05 \Gamma$ and $g = 2 \Gamma$.
    }
    \label{fig:fidelity}
\end{figure}

To characterise the temporal evolution of magnon-qubit entanglement and the fidelity of the Bell-cat state we numerically solve Eq. \eqref{eqn:lindblad_evolution}.
Fig. \ref{fig:fidelity} presents the temporal evolution of $E_{N}$ and $F_{\beta}$ in a noisy environment.
As expected, the fidelity of $\ket{\psi}$ initially starts at an ideal value of unity and gradually decreases, whereas $E_{N}$ attains a permitted maximum value before decreasing due to the decay channels. 


\section{Magnonic Bipartite Cat State}
\label{sec:bipartite}

\begin{figure}[ht]
    \centering
    \includegraphics[width=1\linewidth]{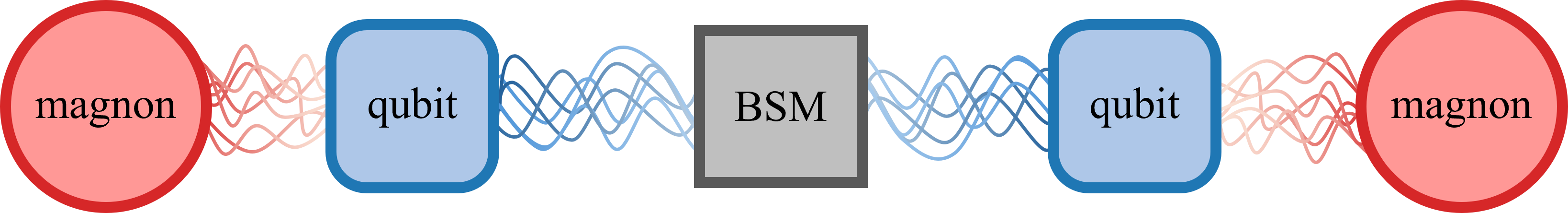}
    \caption{
        (color online) Structure of the combined system after Bell state measurement (BSM).
    }
    \label{fig:combined}
\end{figure}

The above method can be extended to generate bipartite cat states between two spatially separated hybrid systems. 
To achieve this, we connect two identical hybrid magnon-qubit systems described in Sec. \ref{sec:system_hamiltonian} via a common microwave (MW) cavity.
The cavity mediates interactions through its coupling to the respective transmon qubits.
The protocol begins by strongly detuning the transmon qubits from the cavity frequency, thereby suppressing any effective qubit-qubit interaction as well as qubit-mediated interactions between the magnon modes\cite{Song2017,PhysRevLett.123.060502}.
In this regime, magnon-qubit pairs evolve independently under the Lindblad master equation (Eq. \eqref{eqn:lindblad_evolution}) up to a time $\tau$.
Subsequently, the systems are allowed to interact through the MW cavity and a Bell state of the two qubits is generated using the dressed-state phase gate in the presence of qubit dissipation.
Upon performing a projective measurement on the resulting composite system state, the magnons are projected onto bipartite cat states.
A schematic diagram of the protocol is shown in Fig. \ref{fig:combined}.

We denote the states corresponding to the two systems as $\ket{\psi_{j}}$, $j \in \{ 1, 2 \}$ unless stated otherwise.
The entangled states of each pair are allowed to evolve up to $\tau = 1.5 / \Gamma$, at which the entanglement and fidelity reach $E_{N} = 0.89$ and $F_{\beta} = 0.96$ as depicted in Fig. \ref{fig:fidelity}.
The evolution time $\tau$ is chosen such that the generated entanglement between the magnon and qubit remains high while ensuring that $\beta$ has a sufficiently large amplitude, such that the coherent states become nearly orthogonal.
At $\tau$, the flux $\Phi_{M}$ is turned off to stop the interaction between the qubit and the magnon.
At this stage, a displacement operator $\hat{D}(-\beta / 2)$ applied to both the states $\ket{\psi_{1}}$ and $\ket{\psi_{2}}$ results in
\begin{subequations}
\begin{eqnarray}
\label{eqn:combined_states}
\ket{\psi_{1}^{D}} & = & \frac{1}{\sqrt{2}}
\left[ \ket{-\beta / 2}_{m_{1}}\ket{0}_{q_{1}}
+ \ket{\beta / 2}_{m_{1}}\ket{1}_{q_{1}}
\right], \\
\ket{\psi_{2}^{D}} & = & \frac{1}{\sqrt{2}}
\left[
\ket{0}_{q_{2}} \ket{-\beta / 2}_{m_{2}}
+
\ket{1}_{q_{2}} \ket{\beta / 2}_{m_{2}}
\right].
\end{eqnarray}
\end{subequations}
The state of the combined system is given by the tensor product
$\ket{\Psi} = \ket{\psi_{1}^{D}} \otimes \ket{\psi_{2}^{D}}$, such that
\begin{eqnarray}
\label{eqn:combined_tensor}
\ket{\Psi} =
\frac{1}{2}
\big[
& \ket{-\alpha}_1 \ket{0}_1 \ket{0}_2 \ket{-\alpha}_2 \nonumber \\
& + \ket{-\alpha}_1 \ket{0}_1 \ket{1}_2 \ket{\alpha}_2 \nonumber \\
& + \ket{\alpha}_1 \ket{1}_1 \ket{0}_2 \ket{-\alpha}_2 \nonumber \\
& + \ket{\alpha}_1 \ket{1}_1 \ket{1}_2 \ket{\alpha}_2
\big],
\end{eqnarray}
where we have dropped the subscripts for the magnon and the qubit and used $\alpha = \beta/2$.
The combined state can be rewritten in the Bell basis of the qubits $\ket{\psi^{\pm}} = ( \ket{1}_{1} \ket{0}_{2} \pm i \ket{0}_{1} \ket{1}_{2} ) / \sqrt{2} \text{ and } \ket{\phi^{\pm}} = (
\ket{1}_{1} \ket{1}_{2} \pm i \ket{0}_{1} \ket{0}_{2} ) / \sqrt{2}$, which gives us
\begin{align}
\label{eqn:combined_bell_basis}
\ket{\Psi} = \frac{1}{2\sqrt2}
\big[
& \left( \ket{\alpha}_1 \ket{\psi^{-}} \ket{-\alpha}_2 + i \ket{-\alpha}_1 \ket{\psi^{-}} \ket{\alpha}_2 \right) \nonumber \\
& + \left( \ket{\alpha}_1 \ket{\psi^{+}} \ket{-\alpha}_2 - i \ket{-\alpha}_1 \ket{\psi^{+}} \ket{\alpha}_2 \right) \nonumber \\
& + \left( \ket{\alpha}_1 \ket{\phi^{-}} \ket{\alpha}_2 + i \ket{-\alpha}_1 \ket{\phi^{-}} \ket{-\alpha}_2 \right) \nonumber \\
& + \left( \ket{\alpha}_1 \ket{\phi^{+}} \ket{\alpha}_2 - i \ket{-\alpha}_1 \ket{\phi^{+}} \ket{-\alpha}_2 \right)
\big].
\end{align}

\begin{figure}[tp]
    \centering 
    \includegraphics[width=0.98\linewidth]{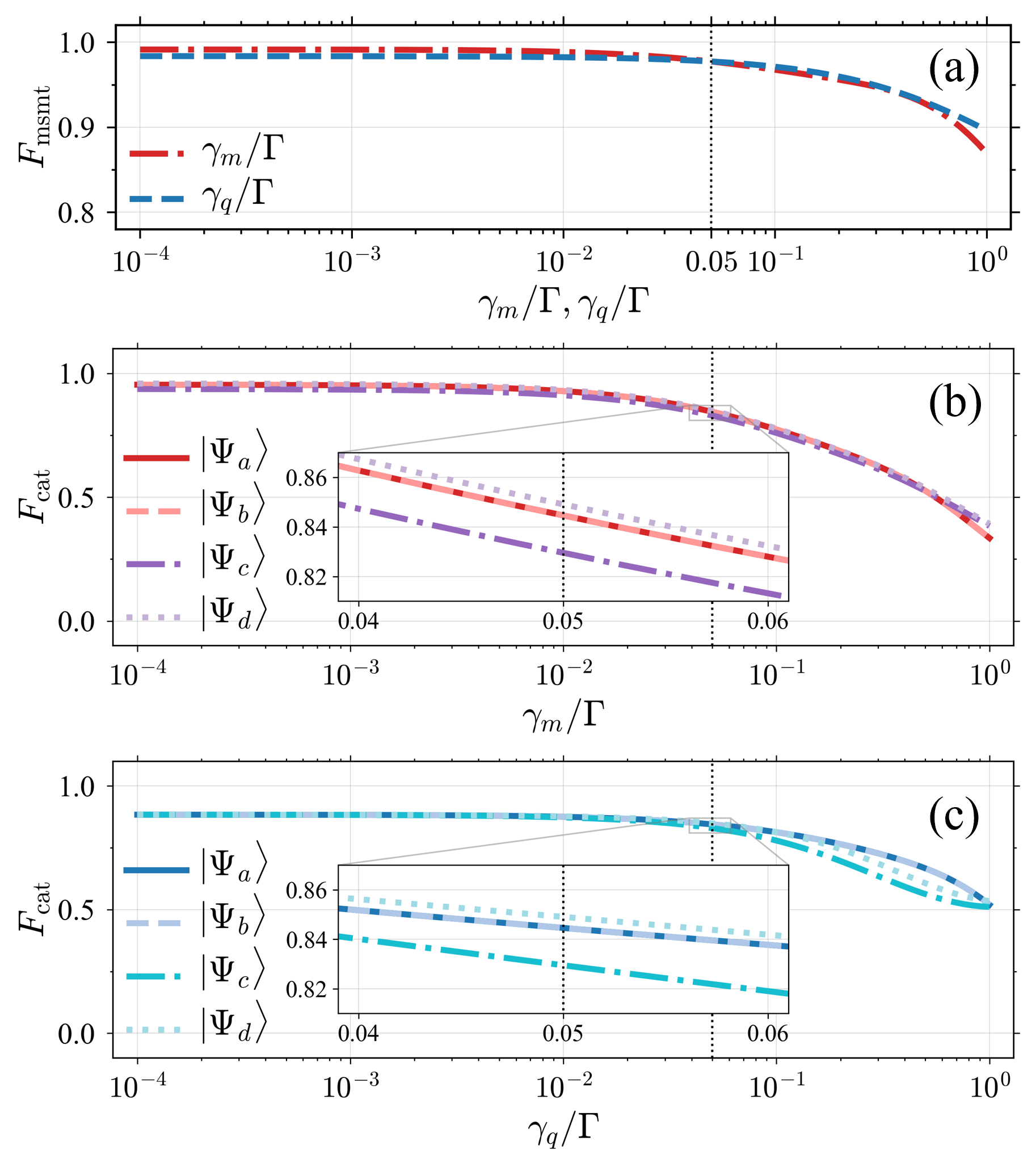}
    \caption{
        (color online)
        (a) Variation of fidelity $F_{\mathrm{msmt}}$ between the measured states obtained using Eqs. \eqref{eqn:phase_unitary} and \eqref{eqn:combined_evolution} for different values of $\gamma_{m}$ and $\gamma_{q}$.
        (b-c) Variation of fidelity $F_{\mathrm{cat}}$ between the measured cat states and the expected cat states for different values of (b) $\gamma_{m}$ and (c) $\gamma_{q}$.
        The dotted lines denote $\gamma_{m}, \gamma_{q} = 0.05 \Gamma$.
        Other parameter are same as Fig. \ref{fig:fidelity}.
    }
    \label{fig:fidelity_dressed}
\end{figure}

As evident from Eq. \eqref{eqn:combined_bell_basis}, a joint Bell-state measurement on the qubits projects the magnonic modes onto one of four bipartite cat states depending on the measurement outcome,
\begin{subequations}
\label{eqn:bipartite_cat_states}
\begin{eqnarray}
\ket{\Psi_{a,b}}
=
\frac{
\ket{\alpha}_1\ket{-\alpha}_2
\pm i \ket{-\alpha}_1\ket{\alpha}_2
}{\sqrt{2}}, \\
\ket{\Psi_{c,d}}
=
\frac{
\ket{\alpha}_1\ket{\alpha}_2
\pm i \ket{-\alpha}_1\ket{-\alpha}_2
}{\sqrt{2}}.
\end{eqnarray}
\end{subequations}

\begin{figure*}[tp]
    \centering
    \includegraphics[width=1\linewidth]{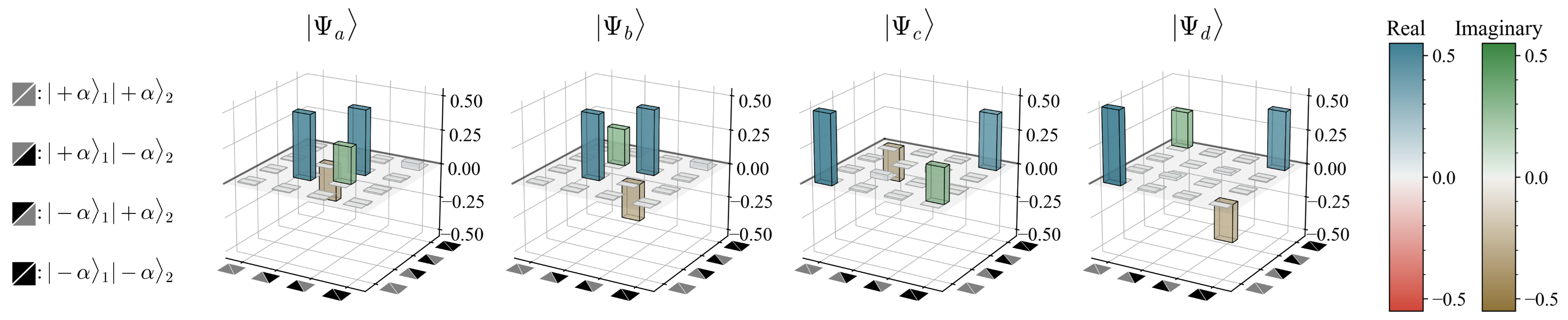}
    \caption{
        (color online) Projection of the generated bipartite magnonic cat states in the two-level coherent subspace.
    }
    \label{fig:histogram_dissipation}
\end{figure*}

To perform a projective Bell-state measurement on the two qubits, they are typically allowed to interact for a fixed amount of time.
This interaction is performed in the dispersive regime, where the qubits are far-detuned from the cavity frequency,
and it is mediated by the virtual exchange of the cavity photons \cite{Nature.460.240}.
However, this exchange results in a two-qubit gate which generates only two of the four Bell states, namely $\ket{\psi^{\pm}}$.
To access all four Bell states, and thereby all four bipartite cat states, we employ a two-qubit dressed-state phase gate protocol implemented via continuous resonant driving of the qubits, as described in detail in Refs. \cite{PhysRevA.110.023726,PhysRevLett.123.060502}.
The unitary matrix corresponding to this dressed-state phase gate protocol is
\begin{equation}
\label{eqn:phase_unitary}
\hat{U} = \frac{1}{\sqrt{2}}
\begin{pmatrix}
1 & 0 & 0 & i \\
0 & 1 & i & 0 \\
0 & i & 1 & 0 \\
i & 0 & 0 & 1
\end{pmatrix}.
\end{equation}
This operator acts on the state in Eq. \eqref{eqn:combined_bell_basis} and maps the Bell basis onto the computational basis as follows:
\begin{subequations}
\begin{eqnarray}
\label{eqn:unitary_matrix}
\hat{U} \ket{\phi^{+}} &= i\ket{0}_1 \ket{0}_2, \quad
\hat{U} \ket{\phi^{-}} &= \ket{1}_1 \ket{1}_2, \\
\hat{U} \ket{\psi^{+}} &= i\ket{0}_1 \ket{1}_2, \quad
\hat{U} \ket{\psi^{-}} &= \ket{1}_1 \ket{0}_2.
\end{eqnarray}
\end{subequations}
A subsequent measurement of the qubits in the computational basis randomly projects the magnons onto one of the four bipartite cat states, $\ket{\Psi_{a, b, c, d}}$.

The entanglement swapping protocol presented here, which transfers entanglement from magnon-qubit pairs to magnon-magnon modes, fundamentally relies on high-fidelity Bell-state measurements, the performance of which depends critically on the dressed-state phase gate.
However, the realisation of two-qubit gates generally suffers from various errors, including coherent and incoherent contributions arising from relaxation and dephasing of individual qubits.
To account for decoherence in the Bell-state measurement implemented through the dressed-state phase gate, we recast $\hat{U}$ as an effective Hamiltonian $\hat{H}_{\hat{U}} = -\hslash \Gamma \hat{\sigma}_{x}^{(1)} \hat{\sigma}_{x}^{(2)}$, which generates the same unitary evolution $\exp{[-i \hat{H}_{\hat{U}} t_{\hat{U}} / \hslash]}$ provided the dynamics is allowed to evolve for a time $t_{\hat{U}} = \pi / (4 \Gamma)$.
To capture the imperfect dynamics, we evolve the density matrix under the master equation including the effects of qubit relaxation and dephasing:
\begin{equation}
\label{eqn:combined_evolution}
\dot{\rho}_{\hat{U}} (t) =
- \frac{i}{\hslash} \left[ \hat{H}_{\hat{U}}, \rho_{\hat{U}} \right]
+ \sum_{j} \gamma_{m} \mathcal{L}\left[ \hat{m}_{j} \right]
+ \sum_{j} \gamma_{q} \mathcal{L}\left[ \hat{c}_{j} \right].
\end{equation}

Fig. \ref{fig:fidelity_dressed}(a) presents the fidelity of the composite state $\ket{\Psi}$ as a function of the magnon and qubit decay rates for an imperfect Bell-state measurement.
Magnon decay primarily diminishes the amplitude ($\ket{\alpha} \rightarrow \ket{e^{-\gamma_{m} t} \alpha}$) of the generated cat states, whereas qubit decay degrades the coherence between the two modes.
To gain further insight into the effects of decay rates on individual cat states, we present the fidelity of the four generated states for each dissipation channel in Figs. \ref{fig:fidelity_dressed}(b-c).
These fidelities exhibit a clear asymmetry under both magnon and qubit decays. 
The states $\ket{\Psi_{a, b}}$ remain nearly degenerate and maintain higher fidelity in contrast to $\ket{\Psi_{c, d}}$, which display splitting.

As the Bell-state measurement is implemented via a dissipative two-qubit gate, different Bell components couple differently to relaxation and dephasing processes \cite{PhysRevB.109.014304,PhysRevB.87.081305}. 
Since each magnonic cat state is obtained by projection onto a specific qubit Bell state (Eq. \eqref{eqn:combined_bell_basis}), we observe different fidelities of the generated states corresponding to each Bell state.
Fig. \ref{fig:histogram_dissipation} shows the reduced density matrix of the generated magnonic bipartite cat state expressed in the joint basis $\left\{ \ket{\alpha}_{1} \ket{\alpha}_{2}, \ket{\alpha}_{1} \ket{-\alpha}_{2}, \ket{-\alpha}_{1} \ket{\alpha}_{2}, \ket{-\alpha}_{1}\ket{-\alpha}_{2} \right\}$.
Ideally, the diagonal elements, representing the populations, and the off-diagonal elements, representing the coherences between the magnon modes, reach an amplitude of 0.5. However, due to the presence of dissipations, the off-diagonal elements are visibly suppressed.

\begin{figure*}[tp]
    \centering 
    \includegraphics[width=0.98\linewidth]{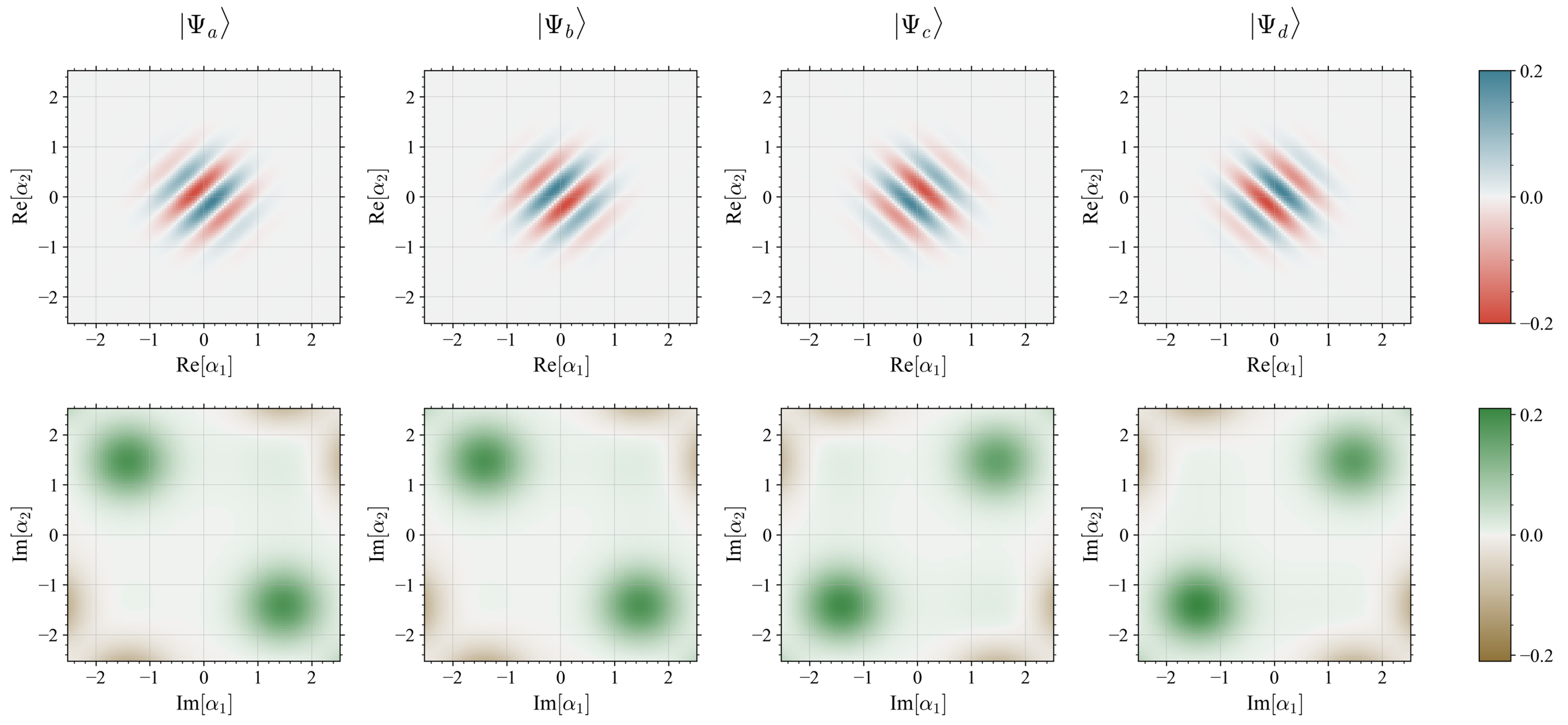}
    \caption{
        (color online) Wigner distributions of the generated bipartite cat states reconstructed numerically using Eq. \eqref{eqn:wigner_joint}.
        The top row depicts the real-real slice of the distribution whereas the bottom row depicts its imaginary-imaginary slice.
    }
    \label{fig:wigner_plots}
\end{figure*}


\section{Joint State Tomography}
\label{sec:tomography}
To verify the prepared state, we employ joint parity-based quantum state tomography, where we compute the joint Wigner function of the bipartite cat state.
This function provides a complete representation of the quantum state as a quasi-probability distribution of bosonic modes in phase space. 
For a bipartite continuous-variable system, the joint Wigner function is defined over a four-dimensional phase space, and captures the correlations between the two modes \cite{Science.352.1087}.

Several techniques have been developed to quantify the joint Wigner function, including homodyne \cite{PhysRevA.94.052327} and heterodyne \cite{RevModPhys.81.299, Chapman:22} detection. 
However, these methods are usually experimentally demanding and may introduce back-action that perturbs the photon state.
A more direct and experimentally efficient alternative is the high-fidelity measurement of photon number parity, which provides direct access to the Wigner distribution \cite{PhysRevLett.89.200402, doi:10.1126/science.1243289}.
It has been shown \cite{PhysRevLett.78.2547} that this distribution can be reconstructed through displaced parity measurements in two steps: (i) applying a displacement operator, and (ii) measuring the expectation value of the displaced parity operator.
This follows the protocol employed in \cite{Science.352.1087} for the case of a two-cavity system, where the joint Wigner function is defined as
\begin{eqnarray}
\label{eqn:wigner_joint}
W(\alpha_{1}, \alpha_{2}) & = & \frac{4}{\pi^{2}} \langle \hat{\Pi}_{J}(\alpha_{1}, \alpha_{2}) \rangle \nonumber \\
& = & \frac{4}{\pi^{2}}\mathrm{Tr} \left[ \rho_{\mathrm{cat}} \hat{D}_{1} \hat{D}_{2} \hat{\Pi}_{J} \hat{D}_{2}^{\dagger} \hat{D}_{1}^{\dagger} \right],
\end{eqnarray}
where $\hat{\Pi}_{J} = \hat{\Pi}_{1} \hat{\Pi}_{2}$ denotes the joint photon-number parity operator, with $\hat{\Pi}_{1} = e^{i\pi\hat{m_{1}}^{\dagger}\hat{m_{1}}}$ and $\hat{\Pi}_{2} = e^{i\pi\hat{m_{2}}^{\dagger}\hat{m_{2}}}$ being the parity operators associated with each mode,
and $\hat{D}_{j} = \hat{D} (\alpha_{j})$, the corresponding displacement operators with coherent amplitudes $\alpha_{j}$ with $\rho_{\mathrm{cat}} $ as the density matrix of the cat states.

It should be noted that while parity measurements have been proposed as an entanglement generation mechanism in magnonic systems \cite{APLQuantum.1.026103}, in the present scheme, they are employed solely for Wigner-function reconstruction, following the successful generation of the bipartite magnonic cat states through projective Bell-state measurement. 
Here, the spatially separated magnon modes are assumed to be dispersively coupled to a three-level transmon, which serves as an ancilla for measurement.
The numerically reconstructed states are shown in Fig. \ref{fig:wigner_plots}.
Since the resultant Wigner function is defined over a four-dimensional phase space, we present two-dimensional cross-sections for visualisation, specifically in the $\mathrm{Re}[\alpha_{1}]$-$\mathrm{Re}[\alpha_{2}]$ plane and the $\mathrm{Im}[\alpha_{1}]$-$\mathrm{Im}[\alpha_{2}]$ plane.
The former exhibits two positive-valued peaks, reflecting the classical mixture of the two coherent components, $\ket{\alpha}$ and $\ket{-\alpha}$.
In contrast, the latter displays interference fringes, indicating a signature of quantum coherence and the nonclassical nature of the bipartite cat states.
For the even and odd bipartite cat states of each pair ($\{\ket{\Psi_{a}}, \ket{\Psi_{b}}\}$ and $\{\ket{\Psi_{c}}, \ket{\Psi_{d}}\}$), the parity symmetry leads to opposite sets of peaks in the interference pattern, as can be seen in the top panel of Fig. \ref{fig:wigner_plots}.

To verify the reconstructed Wigner distributions, we analytically derive the corresponding Wigner functions for each bipartite cat state and compare them with the reconstructed results {(refer to Appendix \ref{app:wigner} for the corresponding expressions)}.
These analytical expressions provide information about (a) the positions of the coherent components and (b) the spacing and direction of the interference fringes.
In the $\mathrm{Re}[\alpha_{1}]$-$\mathrm{Re}[\alpha_{2}]$ plane, the quantities $\alpha^{*} \alpha_{1}$ and $\alpha^{*} \alpha_{2}$ are purely imaginary, which results in the appearance of interference fringes in the Wigner distributions owing to the oscillatory terms, $\sin{\left( 4 \mathrm{Im} \left[ \alpha^{*} \left( \alpha_{1} \pm \alpha_{2} \right) \right] \right)}$.
The directions of these fringes are determined by the signs between $\alpha_{1}$ and $\alpha_{2}$, while $\alpha$ controls their spacing.
On the other hand, in the $\mathrm{Im}[\alpha_{1}]$-$\mathrm{Im}[\alpha_{2}]$ plane, the quantities $\alpha^{*} \alpha_{1}$ and $\alpha^{*} \alpha_{2}$ are purely real, causing the sinusoidal terms to vanish.
The remaining Gaussian terms, $\cosh{\left( 4 \mathrm{Re} \left[ \alpha^{*} \left( \alpha_{1} \pm \alpha_{2} \right) \right] \right)}$, result in two peaks corresponding to the coherent amplitudes.

Furthermore, the structure of the Wigner distribution is strongly influenced by the relative phase of the bipartite cat states $\ket{\Psi_{a, b, c, d}}$.
For instance, if we consider the state $\frac{1}{2} ( \ket{\alpha, \alpha} + \ket{-\alpha, -\alpha} )$ instead of $\ket{\Psi_{c}}$,
we observe an oscillatory dependence on the real part of $\alpha^{*} \left( \alpha_{1} \pm \alpha_{2} \right)$ instead of its imaginary part, that is,
$W \propto \cos{\left( 4 \mathrm{Re} \left[ \alpha^{*} \left( \alpha_{1} \pm \alpha_{2} \right) \right] \right)}$ in the $\mathrm{Im}[\alpha_{1}]$-$\mathrm{Im}[\alpha_{2}]$ plane.
This cosine dependence also gives rise to a central peak in addition to the two displaced Gaussian peaks associated with the coherent-state components.
However, by introducing the relative phase of $\pi/2$ in $\ket{\Psi_{c}}$, we obtain $W \propto \sin{\left( 4 \mathrm{Im} \left[ \alpha^{*} \left( \alpha_{1} \pm \alpha_{2} \right) \right] \right)}$ in the $\mathrm{Re}[\alpha_{1}]$-$\mathrm{Re}[\alpha_{2}]$ plane, without any peaks at the origin.


\section{Conclusion}
\label{sec:conclusion}
In conclusion, we propose a scheme for the remote generation of bipartite magnonic cat states via entanglement swapping between two spatially separated magnon-superconducting qubit hybrid systems.
A magnetised YIG sphere, directly coupled to a flux-tunable transmon qubit through a modulated magnetic flux, enables the formation of a strongly entangled magnon-qubit Bell-cat state.
Subsequent projective Bell-state measurements on the qubits entangle the magnon modes by swapping of entanglement from magnon-qubit to magnon-magnon modes.
Furthermore, the bipartite cat states are shown to retain high entanglement and fidelity even in the presence of dissipation arising from decay of magnon modes, qubit dephasing and relaxation, as well as imperfections in projective measurements.
Finally, we reconstruct the reduced density matrix of the generated state through quantum state tomography of the joint magnonic system. 
By extending high-fidelity parity measurement techniques to magnonic modes, the Wigner distributions corresponding to all four generated states are reconstructed via joint displaced parity measurements mediated by a dispersively coupled ancilla transmon.

\section*{Acknowledgement}
U.D. gratefully acknowledges a research fellowship from MoE, Government of India. R.N. and A.K.S. acknowledges the grant from the MoE, Government of India (Grant No. MoE-STARS/STARS-2/2023-0161).

\appendix


\begin{figure*}[tp]
    \centering 
    \includegraphics[width=0.98\linewidth]{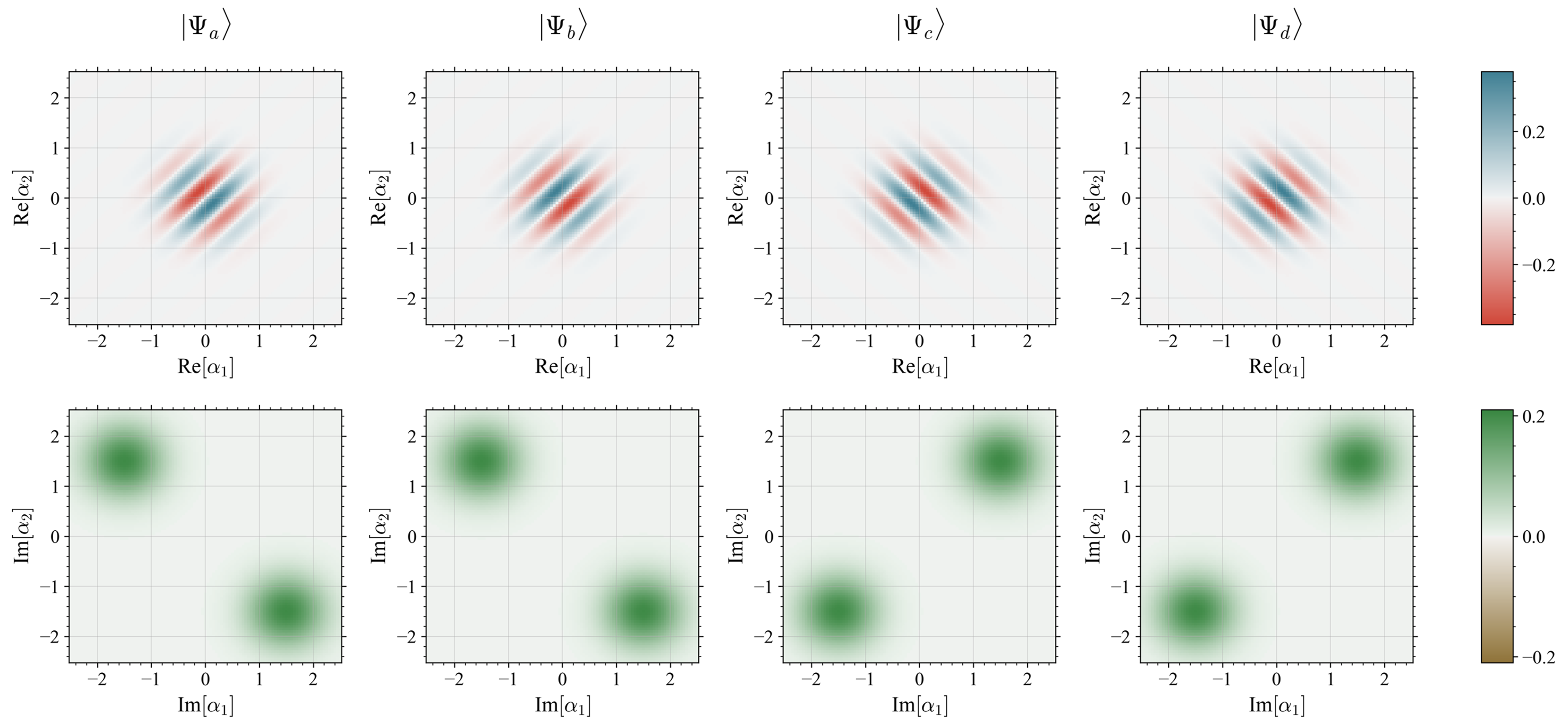}
    \caption{
        (color online) Analytically obtained Wigner functions for the four bipartite cat states using the expressions given in Eqs. \eqref{eqn:wigner_expressions}. Same ordering as in Fig. \ref{fig:wigner_plots}.
}.
    \label{fig:wigner_plots_analytical}
\end{figure*}

\section{Magnon-Qubit Hamiltonian}
\label{app:hamiltonian}

The magnetic moment produced by the stray magnetic field from the YIG sphere results in the complete Hamiltonian
\begin{eqnarray}
\label{eqn:hamiltonian_complete}
\hat{H} & = & \hslash \omega_{m} \hat{m}^{\dagger} \hat{m} + 4 E_{c} \hat{n}^{2} \nonumber \\
&& - \bigg[ E_{JM}^{\mathrm{max}} \cos{\left( \frac{\pi \left( \Phi_{M} + \Phi(\Delta \mu) \right)}{\Phi_{0}} \right)} \nonumber \\
&& + E_{JT}^{\mathrm{max}} \cos{\left( \frac{\pi \Phi_{T}}{\Phi_{0}} \right)} \bigg] \cos{\hat{\delta}} \nonumber \\
& = & \hslash \omega_{m} \hat{m}^{\dagger} \hat{m} + 4 E_{c} \hat{n}^{2} \nonumber \\
&& - \bigg[ E_{JM}^{\mathrm{max}} \cos{\left( \frac{\pi \Phi_{M}}{\Phi_{0}} \right)} \cos{\left( \frac{\pi \Phi(\Delta \mu)}{\Phi_{0}} \right)} \nonumber \\
&& - E_{JM}^{\mathrm{max}} \sin{\left( \frac{\pi \Phi_{M}}{\Phi_{0}} \right)} \sin{\left( \frac{\pi \Phi(\Delta \mu)}{\Phi_{0}} \right)} \nonumber \\
&& + E_{JT}^{\mathrm{max}} \cos{\left( \frac{\pi \Phi_{T}}{\Phi_{0}} \right)} \bigg] \cos{\hat{\delta}}.
\end{eqnarray}

Considering $\Phi(\Delta \mu) \ll \Phi_{0}$, we obtain
\begin{equation}
\hat{H}_{\mathrm{tot}} = \hslash \omega_{m} \hat{m}^{\dagger} \hat{m} + \hat{H}_{\mathrm{int}} + \hat{H}_{TM}^{(0)},
\end{equation}
where $\hat{H}_{\mathrm{int}} = E_{JM}^{\mathrm{max}}\sin{\left( \frac{\pi \Phi_{M}}{\Phi_{0}} \right) \Phi(\Delta \mu)} \cos{\hat{\delta}}$ is the effective magnon-SQUID interaction Hamiltonian.
The coupling between the magnon and the transmon
can be switched off by turning the flux $\Phi_{M}$ off.

In terms of the bosonic creation and annihilation operators, we can write
\begin{subequations}
\begin{eqnarray}
\hat{n} & = \frac{i}{\sqrt{2}} \left( \frac{A}{8 E_{c}} \right)^{\frac{1}{4}} \left( \hat{c}^{\dagger} - \hat{c} \right), \\
\label{delta_expanded}
\hat{\delta} & = \frac{1}{\sqrt{2}} \left( \frac{8 E_{c}}{A} \right)^{\frac{1}{4}} \left( \hat{c}^{\dagger} + \hat{c} \right),
\end{eqnarray}
\end{subequations}
where $A = E_{JM}^{\mathrm{max}} \cos{\left( \pi \Phi_{M} / \Phi_{0} \right)}+ E_{JT}^{\mathrm{max}} \cos{\left( \pi \Phi_{T} / \Phi_{0} \right)}$.
Substituting these in Eq. \eqref{eqn:hamiltonian_squids}, we obtain

\begin{equation}
\hat{H}_{TM}^{(0)} = {\hslash \omega_{q}} \hat{c}^{\dagger} \hat{c} - \frac{E_{c}}{2} \hat{c}^{\dagger} \hat{c}^{\dagger} \hat{c} \hat{c},
\end{equation}
where $\omega_{q} =(\sqrt{8 E_{c} A} - E_{c})$ is the transmon frequency.

For the induced flux, we have
\begin{equation}
\Phi(\Delta \mu) = - \frac{\mu_{0} \mu_{\mathrm{ZPF}}}{4 \Phi_{0} d_{\text{min}}} \left( \hat{m}^{\dagger} + \hat{m} \right),
\end{equation}
where the symbols are defined in the main text.
Under the small-phase approximation, the cosine term is expanded as $\cos{\hat{\delta}} = 1 - \hat{\delta}^{2} / 2$, with higher-order terms neglected. As the constant term only gives rise to an overall energy shift, only the quadratic term is retained, yielding the interaction Hamiltonian
\begin{equation}
\hat{H}_{\text{int}} = \frac{E_{JM}^{\mathrm{max}}\mu_{0} \mu_{\mathrm{ZPF}}}{4 \Phi_{0} d_{\text{min}}} \sin{\left( \frac{\pi \Phi_{M}}{\Phi_{0}} \right)} (\hat{m}^{\dagger} + \hat{m}) \frac{\hat{\delta}^{2}}{2}.
\label{interaction hamiltonian delta}
\end{equation}

Using Eq. \eqref{delta_expanded} and ignoring the highly oscillating terms under RWA, we get
\begin{eqnarray}
\hat{H}_{\text{int}} & = & \frac{E_{JM}^{\mathrm{max}} \mu_{0} \mu_{\text{ZPF}}}{4 \Phi_{0} d_{\text{min}}} \sin{\left( \frac{\pi \Phi_{M}}{\Phi_{0}} \right)} \left( \hat{m}^{\dagger} + \hat{m} \right) \left( \frac{2 E_{c}}{A} \right)^{\frac{1}{2}} \hat{c}^{\dagger} \hat{c} \nonumber \\
& = & \tilde{g} \left( \hat{m}^{\dagger} + \hat{m} \right) \hat{c}^{\dagger} \hat{c},
\end{eqnarray}
where we have defined the coupling strength as
\begin{equation}
\tilde{g} = \frac{E_{JM}^{\mathrm{max}} \mu_{0} \mu_{\text{ZPF}} \sqrt{2 E_{c}}}{4 \hslash \Phi_{0} d_{\text{min}} \sqrt{A}} \sin{\left( \frac{\pi \Phi_{M}}{\Phi_{0}} \right)}.
\end{equation}


\section{Parametric Flux Modulation}
\label{app:modulation}
Substituting $\Phi_{M} = \Phi_{\mathrm{ac}} \cos{(\omega_{\mathrm{ac}} t)}$ in Eq. \eqref{eqn:coupling_strength}, where $\pi \Phi_{\mathrm{ac}} \ll \Phi_{0}$, we obtain
\begin{equation}
\tilde{g} \approx \frac{E_{JM}^{\mathrm{max}} \Phi_{\mathrm{ac}} \mu_{0} \mu_{\text{ZPF}} \sqrt{2 E_{c}}}{4 \hslash \Phi_{0} d_{\text{min}} \sqrt{A_{0}}} \cos{(\omega_{\textrm{ac} t})},
\end{equation}
where $A_{0} = E_{JM}^{\mathrm{max}} + E_{JT}^{\mathrm{max}} \cos{\left( \pi \Phi_{T} / \Phi_{0} \right)}$.
We then rewrite the interaction Hamiltonian in the rotating frame of the magnon mode as
\begin{eqnarray}
\hat{\tilde{H}}_{\mathrm{int}} & = & \frac{E_{JM}^{\mathrm{max}} \Phi_{\mathrm{ac}} \mu_{0} \mu_{\text{ZPF}} \sqrt{2 E_{c}}}{8 \hslash \Phi_{0} d_{\text{min}} \sqrt{A_{0}}} \bigg( \hat{m}^{\dagger} e^{-i (\omega_{\mathrm{ac}} - \omega_{m}) t} \nonumber \\
&& + \hat{m} ^{\dagger} e^{i (\omega_{\mathrm{ac}} + \omega_{m}) t} + \hat{m} e^{-i (\omega_{\mathrm{ac}} + \omega_{m}) t} \nonumber \\
&& + \hat{m} e^{i (\omega_{\mathrm{ac}} - \omega_{m}) t} \bigg) \hat{c}^{\dagger} \hat{c}
\end{eqnarray}

Considering resonant modulation, $\Delta = 0 $ i.e, $\omega_{\mathrm{ac}} = \omega_{m}$, and neglecting the fast-oscillating terms under RWA, we obtain the effective interaction Hamiltonian $\hat{\tilde{H}}_{\mathrm{int}} = g (\hat{m}^{\dagger} + \hat{m}) \hat{c}^{\dagger} \hat{c}$, where $g = 2 \Gamma$ with
\begin{eqnarray}
\Gamma = \frac{E_{JM}^{\mathrm{max}} \Phi_{\mathrm{ac}} \mu_{0} \mu_{\text{ZPF}} \sqrt{2 E_{c}}}{16 \hslash \Phi_{0} d_{\text{min}} \sqrt{A_{0}}}.
\end{eqnarray}


\section{Analytical Derivation for Wigner Function}
\label{app:wigner}

To validate the numerically reconstructed joint Wigner distributions in Fig. \ref{fig:wigner_plots}, we derive the corresponding analytical expressions for all four bipartite cat states.
In what follows, we demonstrate the derivation for the state $\ket{\Psi_{c}}$.
We first decompose the Wigner function in Eq. \eqref{eqn:wigner_joint} for $\ket{\Psi_{c}}$ into two distinct parts
\begin{equation}
\label{eqn:wigner_parts}
W_{c} = W_{\mathrm{I}} + W_{\mathrm{II}},
\end{equation}
where we define the individual terms as
\begin{subequations}
    \begin{eqnarray}
        \label{eqn:wigner_part_1}
        W_{\mathrm{I}} & = & \frac{4}{\pi^{2}} \mathrm{Tr} \left[ \rho_{\mathrm{I}} \hat{D}_{1} \hat{D}_{2} \hat{\Pi}_{1} \hat{\Pi}_{2} \hat{D}_{2}^{\dagger} \hat{D}_{1}^{\dagger} \right]  \\
        \label{eqn:wigner_part_2}
        W_{\mathrm{II}} & = & \frac{4}{\pi^{2}} \mathrm{Tr} \left[\rho_{\mathrm{II}} \hat{D}_{1} \hat{D}_{2} \hat{\Pi}_{1} \hat{\Pi}_{2} \hat{D}_{2}^{\dagger} \hat{D}_{1}^{\dagger} \right], 
    \end{eqnarray}
\end{subequations}
such that $\rho_{c} = \rho_{\mathrm{I}} + \rho_{\mathrm{II}}$ is the density matrix of the state $\ket{\Psi_{c}}$ with $\rho_{\mathrm{I}} = \frac{1}{2} (\ket{\alpha, \alpha} \bra{\alpha, \alpha} + \ket{-\alpha, -\alpha}\bra{-\alpha, -\alpha})$ and $\rho_{\mathrm{II}} = \frac{i}{2}(\ket{-\alpha,-\alpha} \bra{\alpha, \alpha} - \ket{\alpha, \alpha} \bra{-\alpha, -\alpha})$.

While $W_{\mathrm{I}}$ captures the classical mixture of the two coherent components, $W_{\mathrm{II}}$ carries signatures of quantum coherences and the nonclassical nature of the bipartite cat states.

We now write Eq. \eqref{eqn:wigner_part_1} explicitly as
\begin{align}
    W_{\mathrm{I}} & = \frac{2}{\pi^{2}} \bigg[ \bra{\alpha, \alpha} \hat{D}_{1} \hat{D}_{2} \hat{\Pi}_{1} \hat{\Pi}_{2} \hat{D}_{2}^{\dagger} \hat{D}_{1}^{\dagger} \ket{\alpha, \alpha} \nonumber \\
    & + \bra{-\alpha, -\alpha} \hat{D}_{1} \hat{D}_{2} \hat{\Pi}_{1} \hat{\Pi}_{2} \hat{D}_{2}^{\dagger} \hat{D}_{1}^{\dagger} \ket{-\alpha, -\alpha} \bigg].
\end{align}

Since the modes are separable, this factorises into
\begin{align}
W_{\mathrm{I}} & = \frac{2}{\pi^{2}} \bigg[ \prod_{j=1}^{2} \bra{\alpha} \hat{D}_{j} \hat{\Pi}_{j} \hat{D}_{j}^{\dagger} \ket{\alpha} \nonumber \\
& + \prod_{j=1}^{2} \bra{-\alpha} \hat{D}_{j} \hat{\Pi}_{j} \hat{D}_{j}^{\dagger} \ket{-\alpha} \bigg].
\end{align}

Focusing on each mode and using the displacement-operator identity together with the action of the parity operator, we obtain
\begin{align}
    \bra{\alpha} \hat{D}_{j} \hat{\Pi}_{j} \hat{D}_{j}^{\dagger} \ket{\alpha} & = e^{\alpha \alpha^{*}_{j} - \alpha^{*} \alpha_{j}} \langle \alpha \vert - \alpha + 2 \alpha_{j} \rangle \nonumber \\
    & = e^{\alpha \alpha^{*}_{j} - \alpha^{*} \alpha_{j}} e^{-2 \left(\vert \alpha \vert^{2} + \vert \alpha_{j} \vert^{2} \right) + \alpha \alpha_{j}^{*} + 3 \alpha^{*} \alpha_{j}} \nonumber \\
    & = e^{-2 \left( \vert \alpha \vert^{2} + \vert \alpha_{j} \vert^{2} \right)} e^{4 \mathrm{Re} \left[ \alpha^{*} \alpha_{j} \right]},
\end{align}
where we use $\langle \alpha_{j} \vert \alpha_{k} \rangle = \exp{\left[ -\frac{1}{2} \left(\vert \alpha_{j} \vert^{2} + \vert \alpha_{k} \vert^{2} \right)+ \alpha_{j}^{*} \alpha_{k} \right]}$.

Thus, we obtain the diagonal contributions as
\begin{align}
    W_{\mathrm{I}} & = \frac{2}{\pi^{2}} e^{-2 \left( \vert 2 \alpha \vert^{2} + \vert \alpha_{1} \vert^{2} +  \vert \alpha_{2} \vert^{2} \right)} \bigg[ e^{4 \mathrm{Re} \left[ \alpha^{*} \left( \alpha_{1} + \alpha_{2} \right) \right]} \nonumber \\
    & + e^{- 4 \mathrm{Re} \left[ \alpha^{*} \left( \alpha_{1} + \alpha_{2} \right) \right]} \bigg] \nonumber \\
    & = \frac{4}{\pi^{2}} e^{-2 \left( \vert 2\alpha \vert^{2} + \vert \alpha_{1} \vert^{2} +  \vert \alpha_{2} \vert^{2} \right)} \cosh{\left( 4 \mathrm{Re} \left[ \alpha^{*} \left( \alpha_{1} + \alpha_{2} \right) \right] \right)}.
\end{align}

In a similar manner, the second term can be written as
\begin{align}
    W_{\mathrm{II}} & = \frac{2i}{\pi^{2}} \bigg[ \prod_{j=1}^{2} \bra{\alpha} \hat{D}_{j} \hat{\Pi}_{j} \hat{D}_{j}^{\dagger} \ket{- \alpha} \nonumber \\
    & - \prod_{j=1}^{2} \bra{- \alpha} \hat{D}_{j} \hat{\Pi}_{j} \hat{D}_{j}^{\dagger} \ket{\alpha} \bigg].
\end{align}

Here, the individual components can be written as
\begin{align}
    \bra{\pm \alpha} \hat{D}_{j} \hat{\Pi}_{j} \hat{D}_{j}^{\dagger} \ket{\mp \alpha} & = e^{\mp \alpha \alpha^{*}_{j} \pm \alpha^{*} \alpha_{j}} \langle \alpha \vert \alpha + 2 \alpha_{j} \rangle \nonumber \\
    & = e^{\mp \alpha \alpha^{*}_{j} \pm \alpha^{*} \alpha_{j}} e^{-2 \vert \alpha_{j} \vert^{2} \mp \alpha \alpha^{*}_{j} \pm \alpha^{*} \alpha_{j}} \nonumber \\
    & = e^{-2 \vert \alpha_{j} \vert^{2}} e^{\pm 4 i \mathrm{Im} \left[ \alpha^{*} \alpha_{j} \right]}.
\end{align}
Thus, we obtain the interference contribution form the second term as
\begin{align}
    W_{\mathrm{II}} & = \frac{2i}{\pi^{2}} e^{-2 \left( \vert \alpha_{1} \vert^{2} + \vert \alpha_{2} \vert^{2} \right)} \bigg[ e^{4 i \mathrm{Im} \left[ \alpha^{*} \left( \alpha_{1} + \alpha_{2} \right) \right]} \nonumber \\
    & - e^{- 4 i \mathrm{Im} \left[ \alpha^{*} \left( \alpha_{1} + \alpha_{2} \right) \right]} \bigg] \nonumber \\
    & = \frac{-4}{\pi^{2}} e^{-2 \left( \vert \alpha_{1} \vert^{2} + \vert \alpha_{2} \vert^{2} \right)} \sin{\left( 4 \mathrm{Im} \left[ \alpha^{*} \left( \alpha_{1} + \alpha_{2} \right) \right] \right)}.
\end{align}
Following a similar procedure, we obtain all the Wigner functions as
\begin{subequations}
\label{eqn:wigner_expressions}
\begin{align}
    W_{a, b} & = \frac{4}{\pi^{2}} e^{-2 \left( \vert \alpha_{1} \vert^{2} + \vert \alpha_{2} \vert^{2} \right)} \bigg[ e^{-4 \vert \alpha \vert^{2}} \nonumber \\
    & \times \cosh{\left( 4 \mathrm{Re} \left[ \alpha^{*} \left( \alpha_{1} - \alpha_{2} \right) \right] \right)} \nonumber \\
    & \pm \sin{\left( 4 \mathrm{Im} \left[ \alpha^{*} \left( \alpha_{1} - \alpha_{2} \right) \right] \right)} \bigg], \\
    W_{c, d} & = \frac{4}{\pi^{2}} e^{-2 \left( \vert \alpha_{1} \vert^{2} + \vert \alpha_{2} \vert^{2} \right)} \bigg[ e^{-4 \vert \alpha \vert^{2}} \nonumber \\
    & \times \cosh{\left( 4 \mathrm{Re} \left[ \alpha^{*} \left( \alpha_{1} + \alpha_{2} \right) \right] \right)} \nonumber \\
    & \mp \sin{\left( 4 \mathrm{Im} \left[ \alpha^{*} \left( \alpha_{1} + \alpha_{2} \right) \right] \right)} \bigg].
\end{align}
\end{subequations}

Fig. \ref{fig:wigner_plots_analytical} plots the analytically obtained Wigner functions for the four bipartite cat states.
It can be seen that the structures of the distributions in Fig. \ref{fig:wigner_plots_analytical} agree with those in Fig. \ref{fig:wigner_plots}.
The small quantitative differences arise because the numerical reconstruction is performed using simulated tomography data that incorporate dissipation and measurement noise, resulting in a partial suppression of the interference fringes and a slight reduction in the amplitudes of the reconstructed distributions compared with the ideal analytical expressions.

\bibliography{ref}

@article{APLQuantum.1.026103,
    author = {Yan, Jia-shun and Jing, Jun},
    title = {Generating magnon Bell states via parity measurement},
    journal = {APL Quantum},
    volume = {1},
    number = {2},
    pages = {026103},
    year = {2024},
    month = {04},
    issn = {2835-0103},
    doi = {10.1063/5.0201228},
    url = {https://doi.org/10.1063/5.0201228}
}

@article{xu2026nonreciprocal,
  title={Nonreciprocal magnon-magnon entanglement in a spinning cavity-magnon system},
  author={Xu, Zhisheng and Li, Mengxue and Sun, Chunfang and Wang, Gangcheng},
  journal={arXiv preprint arXiv:2605.14394},
  year={2026},
  doi = {10.48550/arXiv.2605.14394},
}

@INPROCEEDINGS{9024573,
  author={Do, Hung and Malaney, Robert and Green, Jonathan},
  booktitle={2019 IEEE Globecom Workshops (GC Wkshps)}, 
  title={Teleportation of a Schrödinger's-Cat State via Satellite-Based Quantum Communications}, 
  year={2019},
  volume={},
  number={},
  pages={1-5},
  doi={10.1109/GCWkshps45667.2019.9024573},
}

@article{Golkar_2024,
  doi = {10.1088/1402-4896/ad0d8d},
  url = {https://doi.org/10.1088/1402-4896/ad0d8d},
  year = {2023},
  month = {dec},
  publisher = {IOP Publishing},
  volume = {99},
  number = {1},
  pages = {015101},
  author = {Golkar, S and Ghasemian, E and Setodeh Kheirabady, M and Tavassoly, M K},
  title = {Magnon-magnon entanglement generation between two remote interaction-free optomagnonic systems via optical Bell-state measurement},
  journal = {Physica Scripta},
}

@article{EtehadiAbari:20,
  author = {Najmeh Etehadi Abari and Mohammad Hossein Naderi},
  journal = {J. Opt. Soc. Am. B},
  number = {7},
  pages = {2146--2156},
  publisher = {Optica Publishing Group},
  title = {Generation of the mechanical Schr\"{o}dinger cat state in a hybrid atom-optomechanical system},
  volume = {37},
  month = {Jul},
  year = {2020},
  url = {https://opg.optica.org/josab/abstract.cfm?URI=josab-37-7-2146},
  doi = {10.1364/JOSAB.393352}
}

@article{Sanders_2012,
  doi = {10.1088/1751-8113/45/24/244002},
  url = {https://doi.org/10.1088/1751-8113/45/24/244002},
  year = {2012},
  month = {may},
  publisher = {IOP Publishing},
  volume = {45},
  number = {24},
  pages = {244002},
  author = {Sanders, Barry C},
  title = {Review of entangled coherent states},
  journal = {Journal of Physics A: Mathematical and Theoretical},
}

@ARTICLE{2021NatRM...6.1114P,
       author = {{Pirro}, Philipp and {Vasyuchka}, Vitaliy I. and {Serga}, Alexander A. and {Hillebrands}, Burkard},
        title = "{Advances in coherent magnonics}",
      journal = {Nature Reviews Materials},
         year = 2021,
        month = dec,
       volume = {6},
       number = {12},
        pages = {1114-1135},
          doi = {10.1038/s41578-021-00332-w},
       adsurl = {https://ui.adsabs.harvard.edu/abs/2021NatRM...6.1114P}
}

@article{Kimble2008,
  author = {Kimble, H. J.},
  title = {The Quantum Internet},
  journal = {Nature},
  volume = {453},
  pages = {1023--1030},
  year = {2008},
  doi = {10.1038/nature07127}
}

@article{Deleglise2008,
  author = {Del{\'e}glise, Samuel and Dotsenko, Igor and Sayrin, Cl{\'e}ment and Bernu, Julien and Brune, Michel and Raimond, Jean-Michel and Haroche, Serge},
  title = {Reconstruction of Non-Classical Cavity Field States with Snapshots of Their Decoherence},
  journal = {Nature},
  year = {2008},
  volume = {455},
  number = {7212},
  pages = {510--514},
  doi = {10.1038/nature07288}
}

@article{Nature.535.262,
  author = {Facon, Adrien and Dietsche, Eva-Katharina and Grosso, Dorian and Haroche, Serge and Raimond, Jean-Michel and Brune, Michel and Gleyzes, S{\'e}bastien},
  title = {A Sensitive Electrometer Based on a Rydberg Atom in a Schr{\"o}dinger-Cat State},
  journal = {Nature},
  year = {2016},
  volume = {535},
  number = {7611},
  pages = {262--265},
  doi = {10.1038/nature18327}
}

@article{hacker.13.110,
  title={Deterministic creation of entangled atom--light Schr{\"o}dinger-cat states},
  author={Hacker, Bastian and Welte, Stephan and Daiss, Severin and Shaukat, Armin and Ritter, Stephan and Li, Lin and Rempe, Gerhard},
  journal={Nature Photonics},
  volume={13},
  number={2},
  pages={110--115},
  year={2019},
  publisher={Nature Publishing Group UK London},
  url = {https://doi.org/10.1038/s41566-018-0339-5}
}

@article{Nature.460.240,
  title = {Demonstration of two-qubit algorithms with a superconducting quantum processor},
  author = {L. DiCarlo and J. M. Chow and J. M. Gambetta and Lev S. Bishop and B. R. Johnson and D. I. Schuster and J. Majer and A. Blais and L. Frunzio and S. M. Girvin and R. J. Schoelkopf},
  journal = {Nature},
  year = {2009},
  volume = {460},
  number = {7252},
  pages = {240--244},
  doi = {10.1038/nature08121},
  url = {https://doi.org/10.1038/nature08121},
  issn = {1476-4687}
}

@article{Nature.511.444,
  author = {Sun, L. and Petrenko, A. and Leghtas, Z. and Vlastakis, B. and Kirchmair, G. and Sliwa, K. M. and Narla, A. and Hatridge, M. and Shankar, S. and Blumoff, J. and Frunzio, L. and Mirrahimi, M. and Devoret, M. H. and Schoelkopf, R. J.},
  title = {Tracking photon jumps with repeated quantum non-demolition parity measurements},
  journal = {Nature},
  volume = {511},
  number = {7510},
  pages = {444--448},
  year = {2014},
  doi = {10.1038/nature13436},
  url = {https://doi.org/10.1038/nature13436}
}

@article{hou2016generation,
  title={Generation of macroscopic Schr{\"o}dinger cat state in diamond mechanical resonator},
  author={Hou, Qizhe and Yang, Wanli and Chen, Changyong and Yin, Zhangqi},
  journal={Sci Rep},
  volume={6},
  number={1},
  pages={37542},
  year={2016},
  publisher={Nature Publishing Group UK London},
  url = {https://doi.org/10.1038/srep37542}
}

@article{Vlastakis2015,
  author = {Vlastakis, Brian and Petrenko, Andrei and Ofek, Nissim and Sun, Luyan and Leghtas, Zaki and Sliwa, Katrina and Liu, Yehan and Hatridge, Michael and Blumoff, Jacob and Frunzio, Luigi and Mirrahimi, Mazyar and Jiang, Liang and Devoret, M. H. and Schoelkopf, R. J.},
  title = {Characterizing entanglement of an artificial atom and a cavity cat state with Bell's inequality},
  journal = {Nature Communications},
  volume = {6},
  number = {1},
  pages = {8970},
  year = {2015},
  doi = {10.1038/ncomms9970},
  url = {https://doi.org/10.1038/ncomms9970}
}

@article{Song2017,
  author = {Song, Chao and Zheng, Shi-Biao and Zhang, Pengfei and Xu, Kai and Zhang, Libo and Guo, Qiujiang and Liu, Wuxin and Xu, Da and Deng, Hui and Huang, Keqiang and Zheng, Dongning and Zhu, Xiaobo and Wang, H.},
  title = {Continuous-variable geometric phase and its manipulation for quantum computation in a superconducting circuit},
  journal = {Nature Communications},
  volume = {8},
  number = {1},
  pages = {1061},
  year = {2017},
  doi = {10.1038/s41467-017-01156-5},
  url = {https://doi.org/10.1038/s41467-017-01156-5}
}

@article{Chapman:22,
author = {Joseph C. Chapman and Joseph M. Lukens and Bing Qi and Raphael C. Pooser and Nicholas A. Peters},
journal = {Opt. Express},
number = {9},
pages = {15184--15200},
publisher = {Optica Publishing Group},
title = {Bayesian homodyne and heterodyne tomography},
volume = {30},
month = {Apr},
year = {2022},
url = {https://opg.optica.org/oe/abstract.cfm?URI=oe-30-9-15184},
doi = {10.1364/OE.456597}
}

@article{PhysRep.1153.1,
  author = {Neill Lambert and Eric Giguère and Paul Menczel and Boxi Li and Patrick Hopf and Gerardo Suárez and Marc Gali and Jake Lishman and Rushiraj Gadhvi and Rochisha Agarwal and Asier Galicia and Nathan Shammah and Paul Nation and J.R. Johansson and Shahnawaz Ahmed and Simon Cross and Alexander Pitchford and Franco Nori},
  title = {QuTiP 5: The Quantum Toolbox in Python},
  journal = {Physics Reports},
  volume = {1153},
  pages = {1-62},
  year = {2026},
  issn = {0370-1573},
  doi = {https://doi.org/10.1016/j.physrep.2025.10.001},
  url = {https://www.sciencedirect.com/science/article/pii/S0370157325002704},
}

@article{PhysRev.58.1098,
  title = {Field Dependence of the Intrinsic Domain Magnetization of a Ferromagnet},
  author = {Holstein, T. and Primakoff, H.},
  journal = {Phys. Rev.},
  volume = {58},
  issue = {12},
  pages = {1098--1113},
  numpages = {0},
  year = {1940},
  month = {Dec},
  publisher = {American Physical Society},
  doi = {10.1103/PhysRev.58.1098},
  url = {https://link.aps.org/doi/10.1103/PhysRev.58.1098}
}

@article{PhysRevA.56.4175,
  title = {Preparation of nonclassical states in cavities with a moving mirror},
  author = {Bose, S. and Jacobs, K. and Knight, P. L.},
  journal = {Phys. Rev. A},
  volume = {56},
  issue = {5},
  pages = {4175--4186},
  numpages = {0},
  year = {1997},
  month = {Nov},
  publisher = {American Physical Society},
  doi = {10.1103/PhysRevA.56.4175},
  url = {https://link.aps.org/doi/10.1103/PhysRevA.56.4175}
}

@article{PhysRevA.65.042305,
  title = {Efficient quantum computation using coherent states},
  author = {Jeong, H. and Kim, M. S.},
  journal = {Phys. Rev. A},
  volume = {65},
  issue = {4},
  pages = {042305},
  numpages = {6},
  year = {2002},
  month = {Mar},
  publisher = {American Physical Society},
  doi = {10.1103/PhysRevA.65.042305},
  url = {https://link.aps.org/doi/10.1103/PhysRevA.65.042305}
}

@article{wbp6-y3vd,
  title = {Digital-analog simulations of Schr\"odinger cat states in the Dicke-Ising model},
  author = {Shapiro, Dmitriy S. and Weber, Yannik and Bode, Tim and Wilhelm, Frank K. and Bagrets, Dmitry},
  journal = {Phys. Rev. A},
  volume = {112},
  issue = {4},
  pages = {042412},
  numpages = {15},
  year = {2025},
  month = {Oct},
  publisher = {American Physical Society},
  doi = {10.1103/wbp6-y3vd},
  url = {https://link.aps.org/doi/10.1103/wbp6-y3vd}
}

@article{PhysRevA.78.062319,
  title = {Quantum repeaters using coherent-state communication},
  author = {van Loock, Peter and L\"utkenhaus, Norbert and Munro, W. J. and Nemoto, Kae},
  journal = {Phys. Rev. A},
  volume = {78},
  issue = {6},
  pages = {062319},
  numpages = {10},
  year = {2008},
  month = {Dec},
  publisher = {American Physical Society},
  doi = {10.1103/PhysRevA.78.062319},
  url = {https://link.aps.org/doi/10.1103/PhysRevA.78.062319}
}

@article{PhysRevA.68.042319,
  title = {Quantum computation with optical coherent states},
  author = {Ralph, T. C. and Gilchrist, A. and Milburn, G. J. and Munro, W. J. and Glancy, S.},
  journal = {Phys. Rev. A},
  volume = {68},
  issue = {4},
  pages = {042319},
  numpages = {11},
  year = {2003},
  month = {Oct},
  publisher = {American Physical Society},
  doi = {10.1103/PhysRevA.68.042319},
  url = {https://link.aps.org/doi/10.1103/PhysRevA.68.042319}
}

@article{PhysRevA.64.022313,
  title = {Entangled coherent states: Teleportation and decoherence},
  author = {van Enk, S. J. and Hirota, O.},
  journal = {Phys. Rev. A},
  volume = {64},
  issue = {2},
  pages = {022313},
  numpages = {6},
  year = {2001},
  month = {Jul},
  publisher = {American Physical Society},
  doi = {10.1103/PhysRevA.64.022313},
  url = {https://link.aps.org/doi/10.1103/PhysRevA.64.022313}
}

@article{PhysRevA.60.674,
  title = {Direct measurement of the Wigner function by photon counting},
  author = {Banaszek, K. and Radzewicz, C. and W\'odkiewicz, K. and Krasi\ifmmode \acute{n}\else \'{n}\fi{}ski, J. S.},
  journal = {Phys. Rev. A},
  volume = {60},
  issue = {1},
  pages = {674--677},
  numpages = {0},
  year = {1999},
  month = {Jul},
  publisher = {American Physical Society},
  doi = {10.1103/PhysRevA.60.674},
  url = {https://link.aps.org/doi/10.1103/PhysRevA.60.674}
}

@article{PhysRevA.109.023703,
  title = {Resonant Schr\"odinger cat states in circuit quantum electrodynamics},
  author = {Ayyash, M. and Xu, X. and Mariantoni, M.},
  journal = {Phys. Rev. A},
  volume = {109},
  issue = {2},
  pages = {023703},
  numpages = {14},
  year = {2024},
  month = {Feb},
  publisher = {American Physical Society},
  doi = {10.1103/PhysRevA.109.023703},
  url = {https://link.aps.org/doi/10.1103/PhysRevA.109.023703}
}

@article{PhysRevLett.127.087203,
  title = {Remote Generation of Magnon Schr\"odinger Cat State via Magnon-Photon Entanglement},
  author = {Sun, Feng-Xiao and Zheng, Sha-Sha and Xiao, Yang and Gong, Qihuang and He, Qiongyi and Xia, Ke},
  journal = {Phys. Rev. Lett.},
  volume = {127},
  issue = {8},
  pages = {087203},
  numpages = {6},
  year = {2021},
  month = {Aug},
  publisher = {American Physical Society},
  doi = {10.1103/PhysRevLett.127.087203},
  url = {https://link.aps.org/doi/10.1103/PhysRevLett.127.087203}
}

@article{zhgm-p3ss,
  title = {Magnon cat states in a cavity-magnon-qubit system via two-magnon driving and dissipation},
  author = {Liu, Gang and Li, Gen and Tan, Huatang and Li, Jie},
  journal = {Phys. Rev. A},
  volume = {112},
  issue = {2},
  pages = {023709},
  numpages = {9},
  year = {2025},
  month = {Aug},
  publisher = {American Physical Society},
  doi = {10.1103/zhgm-p3ss},
  url = {https://link.aps.org/doi/10.1103/zhgm-p3ss}
}

@article{PhysRevA.65.032314,
  title = {Computable measure of entanglement},
  author = {Vidal, G. and Werner, R. F.},
  journal = {Phys. Rev. A},
  volume = {65},
  issue = {3},
  pages = {032314},
  numpages = {11},
  year = {2002},
  month = {Feb},
  publisher = {American Physical Society},
  doi = {10.1103/PhysRevA.65.032314},
  url = {https://link.aps.org/doi/10.1103/PhysRevA.65.032314}
}

@article{PhysRevA.67.012105,
  title = {Maximal violation of Bell inequalities using continuous-variable measurements},
  author = {Wenger, J\'er\^ome and Hafezi, Mohammad and Grosshans, Fr\'ed\'eric and Tualle-Brouri, Rosa and Grangier, Philippe},
  journal = {Phys. Rev. A},
  volume = {67},
  issue = {1},
  pages = {012105},
  numpages = {7},
  year = {2003},
  month = {Jan},
  publisher = {American Physical Society},
  doi = {10.1103/PhysRevA.67.012105},
  url = {https://link.aps.org/doi/10.1103/PhysRevA.67.012105}
}

@article{PhysRevA.110.013711,
  title = {Robust generation of a magnonic cat state via a superconducting flux qubit},
  author = {Hou, Yu-Bo and Hei, Xin-Lei and Pan, Xue-Feng and Xie, Ji-Kun and Ren, Ya-Long and Ma, Sheng-Li and Li, Fu-Li and Li, Peng-Bo},
  journal = {Phys. Rev. A},
  volume = {110},
  issue = {1},
  pages = {013711},
  numpages = {9},
  year = {2024},
  month = {Jul},
  publisher = {American Physical Society},
  doi = {10.1103/PhysRevA.110.013711},
  url = {https://link.aps.org/doi/10.1103/PhysRevA.110.013711}
}

@article{PhysRevA.110.023726,
  title = {Generation of a bipartite mechanical cat state by performing projective Bell-state measurement on a pair of superconducting qubits},
  author = {Nongthombam, Roson and Dewan, Urmimala and Sarma, Amarendra K.},
  journal = {Phys. Rev. A},
  volume = {110},
  issue = {2},
  pages = {023726},
  numpages = {7},
  year = {2024},
  month = {Aug},
  publisher = {American Physical Society},
  doi = {10.1103/PhysRevA.110.023726},
  url = {https://link.aps.org/doi/10.1103/PhysRevA.110.023726}
}

@article{PhysRevA.112.023709,
  title = {Magnon cat states in a cavity-magnon-qubit system via two-magnon driving and dissipation},
  author = {Liu, Gang and Li, Gen and Tan, Huatang and Li, Jie},
  journal = {Phys. Rev. A},
  volume = {112},
  issue = {2},
  pages = {023709},
  numpages = {9},
  year = {2025},
  month = {Aug},
  publisher = {American Physical Society},
  doi = {10.1103/zhgm-p3ss},
  url = {https://link.aps.org/doi/10.1103/zhgm-p3ss}
}

@article{PhysRevA.100.022343,
  title = {Hybrid architecture for engineering magnonic quantum networks},
  author = {Rusconi, C. C. and Schuetz, M. J. A. and Gieseler, J. and Lukin, M. D. and Romero-Isart, O.},
  journal = {Phys. Rev. A},
  volume = {100},
  issue = {2},
  pages = {022343},
  numpages = {14},
  year = {2019},
  month = {Aug},
  publisher = {American Physical Society},
  doi = {10.1103/PhysRevA.100.022343},
  url = {https://link.aps.org/doi/10.1103/PhysRevA.100.022343}
}

@article{PhysRevA.94.052327,
  title = {Optimized tomography of continuous variable systems using excitation counting},
  author = {Shen, Chao and Heeres, Reinier W. and Reinhold, Philip and Jiang, Luyao and Liu, Yi-Kai and Schoelkopf, Robert J. and Jiang, Liang},
  journal = {Phys. Rev. A},
  volume = {94},
  issue = {5},
  pages = {052327},
  numpages = {17},
  year = {2016},
  month = {Nov},
  publisher = {American Physical Society},
  doi = {10.1103/PhysRevA.94.052327},
  url = {https://link.aps.org/doi/10.1103/PhysRevA.94.052327}
}

@article{PhysRevB.104.224302,
  title = {Magnon-magnon entanglement and its quantification via a microwave cavity},
  author = {Azimi Mousolou, Vahid and Liu, Yuefei and Bergman, Anders and Delin, Anna and Eriksson, Olle and Pereiro, Manuel and Thonig, Danny and Sj\"oqvist, Erik},
  journal = {Phys. Rev. B},
  volume = {104},
  issue = {22},
  pages = {224302},
  numpages = {8},
  year = {2021},
  month = {Dec},
  publisher = {American Physical Society},
  doi = {10.1103/PhysRevB.104.224302},
  url = {https://link.aps.org/doi/10.1103/PhysRevB.104.224302}
}

@article{PhysRevB.105.094422,
  title = {Long-range generation of a magnon-magnon entangled state},
  author = {Ren, Ya-long and Xie, Ji-kun and Li, Xin-ke and Ma, Sheng-li and Li, Fu-li},
  journal = {Phys. Rev. B},
  volume = {105},
  issue = {9},
  pages = {094422},
  numpages = {10},
  year = {2022},
  month = {Mar},
  publisher = {American Physical Society},
  doi = {10.1103/PhysRevB.105.094422},
  url = {https://link.aps.org/doi/10.1103/PhysRevB.105.094422}
}

@article{PhysRevB.87.081305,
  title = {Dark Bell states in tunnel-coupled spin qubits},
  author = {S\'anchez, Rafael and Platero, Gloria},
  journal = {Phys. Rev. B},
  volume = {87},
  issue = {8},
  pages = {081305(R)},
  numpages = {5},
  year = {2013},
  month = {Feb},
  publisher = {American Physical Society},
  doi = {10.1103/PhysRevB.87.081305},
  url = {https://link.aps.org/doi/10.1103/PhysRevB.87.081305}
}

@article{PhysRevB.109.014304,
  title = {Optimal encoding of two dissipative interacting qubits},
  author = {Di Bello, G. and De Filippis, G. and Hamma, A. and Perroni, C. A.},
  journal = {Phys. Rev. B},
  volume = {109},
  issue = {1},
  pages = {014304},
  numpages = {14},
  year = {2024},
  month = {Jan},
  publisher = {American Physical Society},
  doi = {10.1103/PhysRevB.109.014304},
  url = {https://link.aps.org/doi/10.1103/PhysRevB.109.014304}
}

@article{PhysRevApplied.21.044018,
  title = {Magnon cat states induced by photon parametric coupling},
  author = {Liu, Da-Wei and Wu, Ying and Si, Liu-Gang},
  journal = {Phys. Rev. Appl.},
  volume = {21},
  issue = {4},
  pages = {044018},
  numpages = {11},
  year = {2024},
  month = {Apr},
  publisher = {American Physical Society},
  doi = {10.1103/PhysRevApplied.21.044018},
  url = {https://link.aps.org/doi/10.1103/PhysRevApplied.21.044018}
}

@article{PhysRevApplied.6.064007,
  title = {Universal Gate for Fixed-Frequency Qubits via a Tunable Bus},
  author = {McKay, David C. and Filipp, Stefan and Mezzacapo, Antonio and Magesan, Easwar and Chow, Jerry M. and Gambetta, Jay M.},
  journal = {Phys. Rev. Appl.},
  volume = {6},
  issue = {6},
  pages = {064007},
  numpages = {10},
  year = {2016},
  month = {Dec},
  publisher = {American Physical Society},
  doi = {10.1103/PhysRevApplied.6.064007},
  url = {https://link.aps.org/doi/10.1103/PhysRevApplied.6.064007}
}

@article{PhysRevLett.81.5932,
  author = {Briegel, H.-J. and D{\"u}r, W. and Cirac, J. I. and Zoller, P.},
  title = {Quantum Repeaters: The Role of Imperfect Local Operations in Quantum Communication},
  journal = {Physical Review Letters},
  volume = {81},
  pages = {5932--5935},
  year = {1998},
  doi = {10.1103/PhysRevLett.81.5932}
}

@article{PhysRevLett.65.3385,
  title = {Collapse and revival of the state vector in the Jaynes-Cummings model: An example of state preparation by a quantum apparatus},
  author = {Gea-Banacloche, Julio},
  journal = {Phys. Rev. Lett.},
  volume = {65},
  issue = {27},
  pages = {3385--3388},
  numpages = {0},
  year = {1990},
  month = {Dec},
  publisher = {American Physical Society},
  doi = {10.1103/PhysRevLett.65.3385},
  url = {https://link.aps.org/doi/10.1103/PhysRevLett.65.3385}
}

@article{PhysRevLett.77.4887,
  title = {Observing the Progressive Decoherence of the ``Meter'' in a Quantum Measurement},
  author = {Brune, M. and Hagley, E. and Dreyer, J. and Ma\^{\i}tre, X. and Maali, A. and Wunderlich, C. and Raimond, J. M. and Haroche, S.},
  journal = {Phys. Rev. Lett.},
  volume = {77},
  issue = {24},
  pages = {4887--4890},
  numpages = {0},
  year = {1996},
  month = {Dec},
  publisher = {American Physical Society},
  doi = {10.1103/PhysRevLett.77.4887},
  url = {https://link.aps.org/doi/10.1103/PhysRevLett.77.4887}
}

@article{PhysRevLett.100.030503,
  title = {Fault-Tolerant Linear Optical Quantum Computing with Small-Amplitude Coherent States},
  author = {Lund, A. P. and Ralph, T. C. and Haselgrove, H. L.},
  journal = {Phys. Rev. Lett.},
  volume = {100},
  issue = {3},
  pages = {030503},
  numpages = {4},
  year = {2008},
  month = {Jan},
  publisher = {American Physical Society},
  doi = {10.1103/PhysRevLett.100.030503},
  url = {https://link.aps.org/doi/10.1103/PhysRevLett.100.030503}
}

@article{PhysRevLett.107.083601,
  title = {Quantum Metrology with Entangled Coherent States},
  author = {Joo, Jaewoo and Munro, William J. and Spiller, Timothy P.},
  journal = {Phys. Rev. Lett.},
  volume = {107},
  issue = {8},
  pages = {083601},
  numpages = {4},
  year = {2011},
  month = {Aug},
  publisher = {American Physical Society},
  doi = {10.1103/PhysRevLett.107.083601},
  url = {https://link.aps.org/doi/10.1103/PhysRevLett.107.083601}
}

@article{PhysRevLett.89.200402,
  title = {Direct Measurement of the Wigner Function of a One-Photon Fock State in a Cavity},
  author = {Bertet, P. and Auffeves, A. and Maioli, P. and Osnaghi, S. and Meunier, T. and Brune, M. and Raimond, J. M. and Haroche, S.},
  journal = {Phys. Rev. Lett.},
  volume = {89},
  issue = {20},
  pages = {200402},
  numpages = {4},
  year = {2002},
  month = {Oct},
  publisher = {American Physical Society},
  doi = {10.1103/PhysRevLett.89.200402},
  url = {https://link.aps.org/doi/10.1103/PhysRevLett.89.200402}
}

@article{PhysRevLett.78.2547,
  title = {Method for Direct Measurement of the Wigner Function in Cavity QED and Ion Traps},
  author = {Lutterbach, L. G. and Davidovich, L.},
  journal = {Phys. Rev. Lett.},
  volume = {78},
  issue = {13},
  pages = {2547--2550},
  numpages = {0},
  year = {1997},
  month = {Mar},
  publisher = {American Physical Society},
  doi = {10.1103/PhysRevLett.78.2547},
  url = {https://link.aps.org/doi/10.1103/PhysRevLett.78.2547}
}

@article{PhysRevLett.123.060502,
  title = {Deterministic Entanglement Swapping in a Superconducting Circuit},
  author = {Ning, Wen and Huang, Xin-Jie and Han, Pei-Rong and Li, Hekang and Deng, Hui and Yang, Zhen-Biao and Zhong, Zhi-Rong and Xia, Yan and Xu, Kai and Zheng, Dongning and Zheng, Shi-Biao},
  journal = {Phys. Rev. Lett.},
  volume = {123},
  issue = {6},
  pages = {060502},
  numpages = {6},
  year = {2019},
  month = {Aug},
  publisher = {American Physical Society},
  doi = {10.1103/PhysRevLett.123.060502},
  url = {https://link.aps.org/doi/10.1103/PhysRevLett.123.060502}
}

@article{PhysRevLett.129.037205,
  title = {Analog Quantum Control of Magnonic Cat States on a Chip by a Superconducting Qubit},
  author = {Kounalakis, Marios and Bauer, Gerrit E. W. and Blanter, Yaroslav M.},
  journal = {Phys. Rev. Lett.},
  volume = {129},
  issue = {3},
  pages = {037205},
  numpages = {7},
  year = {2022},
  month = {Jul},
  publisher = {American Physical Society},
  doi = {10.1103/PhysRevLett.129.037205},
  url = {https://link.aps.org/doi/10.1103/PhysRevLett.129.037205}
}

@article{PhysRevLett.113.156401,
  title = {Strongly Coupled Magnons and Cavity Microwave Photons},
  author = {Zhang, Xufeng and Zou, Chang-Ling and Jiang, Liang and Tang, Hong X.},
  journal = {Phys. Rev. Lett.},
  volume = {113},
  issue = {15},
  pages = {156401},
  numpages = {5},
  year = {2014},
  month = {Oct},
  publisher = {American Physical Society},
  doi = {10.1103/PhysRevLett.113.156401},
  url = {https://link.aps.org/doi/10.1103/PhysRevLett.113.156401}
}

@article{PhysRevLett.113.083603,
  title = {Hybridizing Ferromagnetic Magnons and Microwave Photons in the Quantum Limit},
  author = {Tabuchi, Yutaka and Ishino, Seiichiro and Ishikawa, Toyofumi and Yamazaki, Rekishu and Usami, Koji and Nakamura, Yasunobu},
  journal = {Phys. Rev. Lett.},
  volume = {113},
  issue = {8},
  pages = {083603},
  numpages = {5},
  year = {2014},
  month = {Aug},
  publisher = {American Physical Society},
  doi = {10.1103/PhysRevLett.113.083603},
  url = {https://link.aps.org/doi/10.1103/PhysRevLett.113.083603}
}

@article{PhysRevLett.121.203601,
  title = {Magnon-Photon-Phonon Entanglement in Cavity Magnomechanics},
  author = {Li, Jie and Zhu, Shi-Yao and Agarwal, G. S.},
  journal = {Phys. Rev. Lett.},
  volume = {121},
  issue = {20},
  pages = {203601},
  numpages = {6},
  year = {2018},
  month = {Nov},
  publisher = {American Physical Society},
  doi = {10.1103/PhysRevLett.121.203601},
  url = {https://link.aps.org/doi/10.1103/PhysRevLett.121.203601}
}

@article{PhysRevLett.125.117701,
  title = {Dissipation-Based Quantum Sensing of Magnons with a Superconducting Qubit},
  author = {Wolski, S. P. and Lachance-Quirion, D. and Tabuchi, Y. and Kono, S. and Noguchi, A. and Usami, K. and Nakamura, Y.},
  journal = {Phys. Rev. Lett.},
  volume = {125},
  issue = {11},
  pages = {117701},
  numpages = {6},
  year = {2020},
  month = {Sep},
  publisher = {American Physical Society},
  doi = {10.1103/PhysRevLett.125.117701},
  url = {https://link.aps.org/doi/10.1103/PhysRevLett.125.117701}
}

@article{PRXQuantum.2.040344,
  title = {Quantum Network with Magnonic and Mechanical Nodes},
  author = {Li, Jie and Wang, Yi-Pu and Wu, Wei-Jiang and Zhu, Shi-Yao and You, J.Q.},
  journal = {PRX Quantum},
  volume = {2},
  issue = {4},
  pages = {040344},
  numpages = {13},
  year = {2021},
  month = {Dec},
  publisher = {American Physical Society},
  doi = {10.1103/PRXQuantum.2.040344},
  url = {https://link.aps.org/doi/10.1103/PRXQuantum.2.040344}
}

@article{RevModPhys.81.299,
  title = {Continuous-variable optical quantum-state tomography},
  author = {Lvovsky, A. I. and Raymer, M. G.},
  journal = {Rev. Mod. Phys.},
  volume = {81},
  issue = {1},
  pages = {299--332},
  numpages = {0},
  year = {2009},
  month = {Mar},
  publisher = {American Physical Society},
  doi = {10.1103/RevModPhys.81.299},
  url = {https://link.aps.org/doi/10.1103/RevModPhys.81.299},
}

@article{doi:10.1126/science.aam9288,
  author = {Wehner, Stephanie and Elkouss, David and Hanson, Ronald},
  title = {Quantum Internet: A Vision for the Road Ahead},
  journal = {Science},
  volume = {362},
  number = {6412},
  year = {2018},
  doi = {10.1126/science.aam9288}
}

@article{SciAdv.8.eabn1778,
  author = {Z. Wang and Z. Bao and Y. Wu and Y. Li and W. Cai and W. Wang and Y. Ma and T. Cai and X. Han and J. Wang and Y. Song and L. Sun and H. Zhang and L. Duan},
  title = {A flying Schrödinger's cat in multipartite entangled states},
  journal = {Sci. Adv.},
  volume = {8},
  number = {10},
  pages = {eabn1778},
  year = {2022},
  doi = {10.1126/sciadv.abn1778},
  url = {https://www.science.org/doi/10.1126/sciadv.abn1778},
}

@article{Science.352.1087,
  title = {A Schrödinger cat living in two boxes},
  author = {Chen Wang  and Yvonne Y. Gao  and Philip Reinhold  and R. W. Heeres  and Nissim Ofek  and Kevin Chou  and Christopher Axline  and Matthew Reagor  and Jacob Blumoff  and K. M. Sliwa  and L. Frunzio  and S. M. Girvin  and Liang Jiang  and M. Mirrahimi  and M. H. Devoret  and R. J. Schoelkopf },
  journal = {Science},
  volume = {352},
  number = {6289},
  pages = {1087-1091},
  year = {2016},
  doi = {10.1126/science.aaf2941},
  URL = {https://www.science.org/doi/abs/10.1126/science.aaf2941},
}

@article{doi:10.1126/science.adf7553,
  author = {Marius Bild  and Matteo Fadel  and Yu Yang  and Uwe von Lüpke  and Phillip Martin  and Alessandro Bruno  and Yiwen Chu },
  title = {Schrödinger cat states of a 16-microgram mechanical oscillator},
  journal = {Science},
  volume = {380},
  number = {6642},
  pages = {274-278},
  year = {2023},
  doi = {10.1126/science.adf7553},
  URL = {https://www.science.org/doi/abs/10.1126/science.adf7553},
}

@article{doi:10.1126/science.1243289,
  author = {Brian Vlastakis  and Gerhard Kirchmair  and Zaki Leghtas  and Simon E. Nigg  and Luigi Frunzio  and S. M. Girvin  and Mazyar Mirrahimi  and M. H. Devoret  and R. J. Schoelkopf },
  title = {Deterministically Encoding Quantum Information Using 100-Photon Schrödinger Cat States},
  journal = {Science},
  volume = {342},
  number = {6158},
  pages = {607-610},
  year = {2013},
  doi = {10.1126/science.1243289},
  URL = {https://www.science.org/doi/abs/10.1126/science.1243289},
}

@article{doi:10.1126/science.aaa3693,
  author = {Yutaka Tabuchi  and Seiichiro Ishino  and Atsushi Noguchi  and Toyofumi Ishikawa  and Rekishu Yamazaki  and Koji Usami  and Yasunobu Nakamura },
  title = {Coherent coupling between a ferromagnetic magnon and a superconducting qubit},
  journal = {Science},
  volume = {349},
  number = {6246},
  pages = {405-408},
  year = {2015},
  doi = {10.1126/science.aaa3693},
  URL = {https://www.science.org/doi/abs/10.1126/science.aaa3693},
}

@article{doi:10.1126/sciadv.1501286,
  author = {Xufeng Zhang  and Chang-Ling Zou  and Liang Jiang  and Hong X. Tang },
  title = {Cavity magnomechanics},
  journal = {Science Advances},
  volume = {2},
  number = {3},
  pages = {e1501286},
  year = {2016},
  doi = {10.1126/sciadv.1501286},
  URL = {https://www.science.org/doi/abs/10.1126/sciadv.1501286},
}

@article{doi:10.1126/science.aaz9236,
  author = {Dany Lachance-Quirion  and Samuel Piotr Wolski  and Yutaka Tabuchi  and Shingo Kono  and Koji Usami  and Yasunobu Nakamura },
  title = {Entanglement-based single-shot detection of a single magnon with a superconducting qubit},
  journal = {Science},
  volume = {367},
  number = {6476},
  pages = {425-428},
  year = {2020},
  doi = {10.1126/science.aaz9236},
  URL = {https://www.science.org/doi/abs/10.1126/science.aaz9236},
}

\end{document}